\documentclass{aa}

\usepackage[varg]{txfonts}

\usepackage{rotating}
\usepackage{xspace}
\usepackage{amssymb}
\usepackage{graphicx}
\usepackage{booktabs}
\usepackage{hyperref}
\usepackage{txfonts}
\usepackage{graphicx, float}
\usepackage{gensymb}
\usepackage{natbib}
\usepackage{epsfig, natbib}

\begin{document}

\title
{Insights from the Outskirts: Chemical and Dynamical Properties in the outer Parts of the Fornax Dwarf Spheroidal Galaxy\thanks{This article is based on observations made with ESO Telescopes at the Paranal Observatory under programme 082.B-0940(A).}}

\author{Benjamin Hendricks\inst{\ref{inst1}} \and Andreas Koch\inst{\ref{inst1}} \and Matthew Walker\inst{\ref{inst2}} \and Christian I. Johnson\inst{\ref{inst3}} \and Jorge Pe\~{n}arrubia\inst{\ref{inst4}} \and Gerard Gilmore\inst{\ref{inst5}}}

\institute{Zentrum f\"ur Astronomie der Universit\"at Heidelberg, Landessternwarte, K\"onigstuhl 12, 69117, Heidelberg, Germany\label{inst1} \and
McWilliams Center for Cosmology, Carnegie Mellon University, 5000 Forbes Ave., Pittsburgh, PA 15213, USA\label{inst2} \and
Harvard-Smithsonian Center for Astrophysics, 60 Garden Street, MS-15, Cambridge, MA 02138, USA\label{inst3} \and
Institute for Astronomy, University of Edinburgh, Royal Observatory, Blackford Hill, Edinburgh EH9 3HJ, UK\label{inst4} \and
Institute of Astronomy, Cambridge University, Madingley Rd, Cambridge CB3 OHA, UK\label{inst5} }

\date{21 July 2014 / 21 August 2014 }

\abstract{We present radial velocities and [Fe/H] abundances for 340 stars in the Fornax dwarf spheroidal from $R\sim16,000$ spectra. The targets have been obtained in the outer parts of the galaxy, a region which has been poorly studied before.
Our sample shows a wide range in [Fe/H], between $-0.5$ and $-3.0$\,dex, in which we detect three subgroups. 
Removal of stars belonging to the most metal-rich population produces a truncated metallicity distribution function that is identical to Sculptor, indicating that these systems have shared a similar early evolution, only that Fornax experienced a late, intense period of star formation (SF).
The derived age-metallicity relation shows a fast increase in $\mathrm{[Fe/H]}$ at early ages, after which the enrichment flattens significantly for stars younger than $\sim8$\,Gyr. Additionally, the data indicate a strong population of stars around $4$\,Gyr, followed by a second rapid enrichment in [Fe/H]. 
A leaky-box chemical enrichment model generally matches the observed relation but does not predict a significant population of young stars nor the strong enrichment at late times. The young population in Fornax may therefore originate from an externally triggered SF event. 
Our dynamical analysis reveals an increasing velocity dispersion with decreasing [Fe/H] from $\sigma_{sys}\approx 7.5\,\mathrm{kms^{-1}}$ to $\geq14,\mathrm{kms^{-1}}$, indicating an outside-in star formation history in a dark matter dominated halo. 
The large velocity dispersion at low metallicities is possibly the result of a non-Gaussian velocity distribution amongst stars older than $\sim8$\,Gyr.
Our sample also includes members from the Fornax GCs H2 and H5. In agreement with past studies we find $\mathrm{[Fe/H]}=-2.04\pm0.04$ and a mean radial velocity $RV=59.36\pm0.31$\,$\mathrm{km\,s^{\rm -1}}$ for H2 and $\mathrm{[Fe/H]}=-2.02\pm0.11$ and $RV=59.39\pm0.44$\,$\mathrm{km\,s^{\rm -1}}$ for H5.
Finally, we test different calibrations of the Calcium Triplet over more than 2\,dex in $\mathrm{[Fe/H]}$ and find best agreement with the calibration equations provided by \citet{Carrera_13}. Overall, we find large complexity in the chemical and dynamical properties, with signatures that additionally vary with galactocentric distance. Detailed knowledge about the properties of stars at all radii is therefore necessary to draw a conclusive picture about the star formation and chemical evolution in Fornax.}


\titlerunning{Chemical and Dynamical Properties in the outer Parts of the Fornax dSph}
\authorrunning{Hendricks et al.}
\maketitle

\section{Introduction} 
\label{chap_01}

Satellite galaxies in the Local Group provide excellent laboratories to study the chemical and dynamical properties within these systems, because they can be dissected on a star-by-star basis. While they are sufficiently close to be approached with medium- and high resolution spectroscopy for detailed chemical analysis of their brightest giants, they are still sufficiently far away to cover a significant fraction of their surface profile with few individual pointings. Thus, important insights can be drawn on their evolutionary pathways and about their chemodynamical complexity. 

Amongst the dwarf galaxies, the Local Group dwarf spheroidal (dSph) galaxies are the smallest, closest and most abundant systems in the Local Universe. These galaxies are specifically simple, because they are not actively forming stars at present, and they do not contain a significant amount of gas (\citealt{Grebel_03}).
Therefore, all the byproducts of stellar evolution within the galaxy can be assumed to be retained in stars, and thus be measurable, unless they are drawn away from the galaxy through galactic winds (e.g., \citealt{Lanfranchi_07}).
The large number of dSphs in the Local Group (we know of, and can perform star-by-star studies on $\sim 30$, e.g., \citealt{McConnachie_12}, \citealt{Weisz_14}) opens the chance to compare their individual characteristics and thus have the possibility to unravel universal parameters that regulate their star formation history (SFH) and chemical enrichment.
Thus, it becomes possible to connect the mechanisms of the satellite galaxies to the more complex stellar systems like our Milky Way (MW) and their influence on each other.

Although dSphs have stellar masses typically $\leq 10^{7} M_{\sun}$, there is a large complexity among their chemical and dynamical properties (e.g., \citealt{Grebel_03}, \citealt{Tolstoy_09}). Recently, \citet{Weisz_14} found significant scatter in the SFH of Local Group dSphs, even if only galaxies of similar mass are compared, which indicates that a diversity of environmental influences must have had an important impact on the evolution of theses systems. Such interactions can have contrary effects: while tidal- and ram pressure stripping can slow down or even quench star formation (SF) in a galaxy (\citealt{Mayer_06}), the accretion of gas or merger events may trigger SF bursts and alter the chemical enrichment history. Given their shallow gravitational potentials compared to larger galaxies, dSphs should be most sensitive to such effects, which makes them important testing grounds to understand the frequency and impact of the aforementioned external influences.

The knowledge about detailed chemodynamical properties in satellite galaxies evolved particularly with the advent of powerful, fiber-fed multi-object spectrographs, which enable us to obtain simultaneously precise velocity information and chemical abundances for a large number of stars.
Therefore, today large samples of more than 50 stars with at least metallicity\footnote{Throughout the remainder of this paper the term \emph{metallicity} and \emph{[Fe/H]} will be used interchangeably.} and velocity measurements exist for all of the more luminous dSphs associated with the MW: Carina (\citealt{Koch_06}, \citealt{Lemasle_12}), Sextans (\citealt{Battaglia_11}), Sculptor (\citealt{Tolstoy_09}), Draco and Ursa Minor (\citealt{Kirby_11c}), Leo\,I and\,II (\citealt{Koch_07a, Koch_07b}), Sagittarius (\citealt{Carretta_10}), and Fornax (\citealt{Pont_04}, \citealt{Battaglia_06}). Note, that \citet{Kirby_11c} provides spectroscopic samples for all these dSphs.

The majority of the abovementioned studies make use of the Calcium Triplet (CaT) absorption lines in the near-infrared as an indicator for $\mathrm{[Fe/H]}$ (\citealt{Armandroff_88},\citealt{Rutledge_97b}), motivated by the fact that the CaT is the strongest feature in near-infrared spectra of late-type giant stars. Thus, it can be analyzed even from low- to medium resolution spectra ($R\leq10,000$) with low signal-to-noise (S/N$\sim 10-20$), where individual iron lines can hardly be used (but see \citealt{Kirby_08} for an alternative approach). Unfortunately, the CaT-[Fe/H] calibration relies on several factors such as $log\,g$, $T_{eff}$, and [Ca/Fe], which limits the validity of empirical calibration equations and makes them uncertain especially at extreme metallicities, where few or no calibrators can be found (e.g., \citealt{Battaglia_08}). 

At a distance of $\sim\!\!147$\,kpc (\citealt{Pietrzynski_09}) Fornax is amongst the most massive dSphs in the Local Group, and besides Sagittarius the only dSph galaxy with its own GC system. 
Recent proper motion studies with both ground-based telescopes (\citealt{Walker_08}, \citealt{Mendez_11}) and the Hubble Space Telescope (\citealt{Dinescu_04}, \citealt{Piatek_07}) agree that the current orbital position of Fornax is close to perigalacticon, which it passed less than 1\,Gyr ago. Most of these studies furthermore predict an orbital period of $\sim6$\,Gyr, which implies that Fornax experienced at least two full orbits around the MW during its evolution. In contrast to these studies, \citet{Mendez_11} derive a significantly larger orbital period of 21\,Gyr paired with an extremely high eccentricity. While the orbital information may play an important role on the evolution of dSphs, the evident discrepancies illustrate the large uncertainty in the orbital properties of Fornax, in particular for large look-back times.

Previous chemical and photometric studies have shown a complex and extended SFH including stars older than $12$\,Gyr until stars as young as $\sim250$\,Myr (\citealt{Stetson_98}, \citealt{de_Boer_12}). Moreover, several features have been identified which support either merger events or an otherwise externally or internally disturbed SFH: cold velocity substructures in the central parts (\citealt{Battaglia_06}), different angular momentum vectors for different metallicity populations (\citealt{Amorisco_12}), stellar over-densities (``shells'') in the field (\citealt{Coleman_04}), a strong radial population gradient with metal-rich stars concentrated closer to the center (\citealt{Battaglia_06}), and the chemical evolution of the $\mathrm{\alpha}$-elements (\citealt{Hendricks_14}).
Although it seems as if Fornax (almost) continuously formed stars during the last $\sim\!13$\,Gyr (\citealt{de_Boer_12}), many questions remained unanswered: did Fornax evolve in relative isolation, or did it experience merger events (\citealt{Coleman_04}, \citealt{Battaglia_06}, \citealt{Yozin_12}, \citealt{Amorisco_12}). There is also discussion about the mixing efficiency within the galaxy and the impact of SF bursts on the interstellar medium (ISM). Should one expect to find local inhomogeneities, caused by few individual supernova explosions (\citealt{Marcolini_08})? Was Fornax able to retain some of the gas initially lost in galactic winds which subsequently was re-accreted to the galaxy and became available for SF (\citealt{Ruiz_13}, \citealt{dErcole_99})? Furthermore, it is not clear whether the MW host galaxy or other environmental influences play an important role in the chemodynamical evolution of Fornax. Has the SFH been influenced by periodical tidal interactions (\citealt{Nichols_12})? Did ram pressure stripping caused by AGN shock shells from the MW in the past trigger SF bursts and simultaneously remove large quantities of its (former) gas reservoir (\citealt{Nayakshin_13})?
Finally, why did Fornax form GCs -- while most other dwarfs did not -- and why are not all of them dissolved yet (\citealt{Penarrubia_09})? Consequently it is not known if and how many stars in the field are in fact stripped from existing GCs, or the remnants of already completely dissolved globulars (\citealt{Larsen_12a}).

Most of these aspects are clearly not problems specific to the Fornax dSph but concern most, if not all satellite systems in the Local Group. Constraining open questions concerning the evolutionary pathway of Fornax, will therefore have a direct implication on our understanding of the nature of dwarf galaxies in general.

Here, the combination of spectroscopic and photometric information is particularly powerful, because -- when combined -- stellar ages can be derived and links between dynamical and chemical properties can help to identify and distinguish different origins of individual sub-populations.
For Fornax, \citet{Battaglia_06} provide $562$ spectra distributed throughout the galaxy. About half of these stars are located within $r_{ell}\leq0.3$, about equivalent to Fornax' core radius (see \citealt{Battaglia_06}). 
Additionally, \citet{Pont_04} provide a sample of $117$ stars from the central area with maximal radii of $\sim0.2$\degree, and \citet{Kirby_11c} analyzed $675$ Fornax field stars within a similarly small radius.
Note, that in addition to radial velocities, \citet{Walker_09} also provide [Fe/H] measurements from Mg absorption features, which however show large systematic variations compared to direct Fe or CaT-measurements and therefore are not suited for direct comparison with other samples or for the estimation of actual [Fe/H].
Several later chemodynamical studies use the Battaglia-sample (\citealt{Coleman_08}, \citealt{Amorisco_12}) or use a central subsample of the same targets for high-resolution follow-up (\citealt{Letarte_10}). Consequently, the outer radii of Fornax are still poorly analyzed despite the fact that the chemical evolution shows clear radial trends within its gravitational potential.
Consequently, a complete picture of the chemical evolution of Fornax is only possible if the chemodynamical characteristics at all radii are known, and when their differences are understood. This is specifically important with regard to possible accretion events, since they most likely leave imprints in the outer parts of a galaxy (e.g., \citealt{Naab_09}, \citealt{Brodie_14}).
Simultaneously, the existing sample of metal-poor ($\mathrm{[Fe.H]} \leq-2.0$) stars in Fornax, which bear the information on early chemical evolution is still limited ($\leq60$ throughout the whole galaxy).

Here, we present a chemodynamical analysis for a large sample of stars in the Fornax dSph obtained at large radii within the galaxy. The sample is intended to obtain insights from the outskirts of Fornax and, in combination with the existing samples, provide a tool to pin-down and understand the chemical and dynamical \emph{differences} within this complex galaxy.

This manuscript is organized as follows. In Section \S2, we summarize our data and describe the CaT-analysis and radial velocity (RV) measurements in detail.
In Section \S3, we put on the test different calibration equations for the CaT and discuss possible systematic differences.
In Section \S4, we determine individual stellar ages and discuss the resulting age-metallicity relation (AMR) and age-RV-relation with respect to the chemical enrichment history of Fornax. Special attention is given to the treatment of statistical and systematic uncertainties in the age-determination.
In Section \S5, we show the metallicity distribution function (MDF) of our sample and investigate different sub-populations.
Section \S6 contains our analysis of radial properties, of both metallicity and stellar ages within our sample.
Finally, in Section \S7 we summarize the properties of our spectroscopic sample and highlight the implications to the evolution of Fornax.

\section{Data}
\label{chap_02}
\begin{figure}[htb]
\begin{center}
\includegraphics[width=0.5\textwidth]{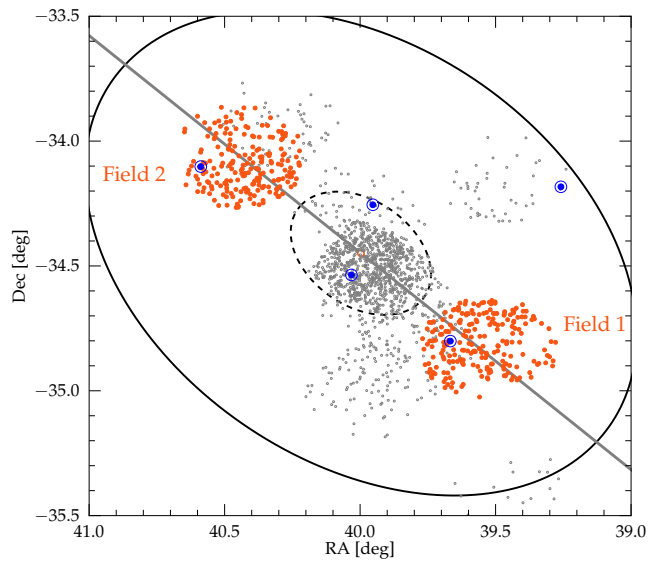}
\caption{Location of our targets (red dots) in the field of Fornax. The samples of previous studies for which RVs and [Fe/H] were obtained (see text) are shown in gray. GCs are shown with blue symbols. To guide the eye, we also indicate the tidal and the core radius of Fornax at $r_{ell}=1.06\degree$ and $r_{ell}=0.29\degree$, respectively.}
\label{fig_field_plot}
\end{center}
\end{figure}

\begin{figure}[htb]
\begin{center}
\includegraphics[width=0.5\textwidth]{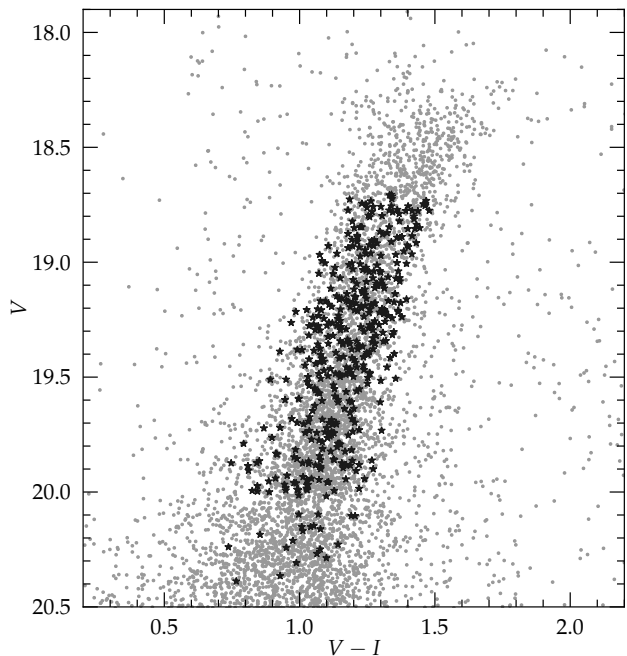}
\caption{Location of our targets (black symbols) on top of the RGB of Fornax drawn from the photometry which we use for target selection. }
\label{fig_CMD}
\end{center}
\end{figure}

The spectra for this study have been obtained in November 2008 with FLAMES at the VLT (programme ID 082.B-0940(A)) , where we used GIRAFFE in MEDUSA high-resolution mode (HR\,21, $R\sim16000$, $8484 - 9001$\,\AA). With a total integration time for each pointing of 8 hours we obtain a typical S/N of $20-50$ per pixel. 
As shown in Figure\,\ref{fig_field_plot} The targets are distributed in two opposite fields along the major axis of the galaxy, aiming specifically for stars in the outer part of Fornax at distances $r_{ell} \approx 0.4$--$0.8 \degree$. 
Our sample contains 431 bona-fide Fornax members and was selected from optical $V$ and $I$ broadband photometry (\citealt{Walker_06}) within a broad selection box around the red giant branch (RGB), spanning down to the horizontal branch and sampling the full color range of the RGB with the intention to equally include the most metal-rich and metal-poor populations as well as the full age range (see Figure\,\ref{fig_CMD}). 

The spectroscopic sample we use in this study is the same as presented in \citet{Hendricks_14}, which emphasized a detailed chemical abundance analysis for several $\mathrm{\alpha}$-elements. Here, we will discuss in detail the reduction and analysis of the dynamical properties and metallicities derived from the CaT, while we point the reader to the aforementioned paper for details about the pre-reduction process of the spectra and the high-resolution chemical abundance analysis to obtain $\mathrm{[Fe/H]}$ from iron absorption features as well as the individual $\mathrm{\alpha}$-elements.

Note that for all but the next Section we will use [Fe/H] as derived from the CaT and not the direct measurements from Fe absorption features. The main reason is that the CaT can be evaluated for spectra at practically all S/N and over the full range of metallicity. In contrast, we obtain [Fe/H] from Fe absorption lines only for a smaller subsample (331 out of 401 with CaT measurements) with higher S/N, which is additionally biased towards metal-rich stars, for which [Fe/H] can be obtained more easily. Several parts of our analysis, however, require an unbiased sample which reflects the actual distribution of chemical enrichment. Such a set can only be provided from CaT measurements, with the additional advantage of being directly comparable to previous studies in Fornax and other dSphs, that are based on CaT metallicities.

\subsection{Radial Velocities and Galaxy Membership}
\label{radial_velocities}
\begin{figure}[htb]
\begin{center}
\includegraphics[width=0.5\textwidth]{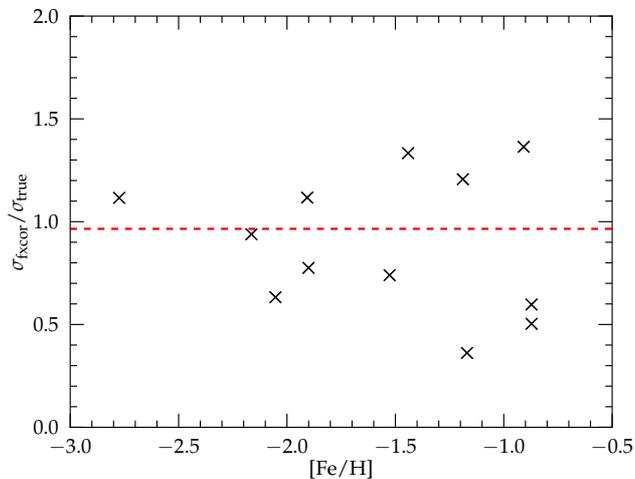}
\caption{Comparison of the mean velocity error from \emph{fxcor} and the standard deviation of multiple measurements for a subset of targets. Almost all ratios fall within 0.5 and 1.5 and are scattered around a mean value of $\sigma_{\mathrm{fxcor}}/ \sigma_{\mathrm{true}}= 0.97\pm0.10$, indicated by the red horizontal line. Note that three stars without [Fe/H] information are not shown here, but have error-ratios of 1.92, 0.83, and 1.04, respectively.}
\label{fxcor_err_fig}
\end{center}
\end{figure}

We determine the line-of-sight radial velocity (RV) for each star via Fourier cross-correlation by comparison to a synthetic CaT template spectrum (\citealt{Kleyna_04}) using \textit{IRAF.fxcor}, which yield typical fitting errors $\leq 1\,\mathrm{km\,s^{\rm -1}}$. 
The evaluation of dynamical properties -- especially the intrinsic velocity dispersion -- fundamentally relies on accurate error estimates for the individual stellar velocities. Hereby the systematic bias gains dramatically in weight as larger the fraction between the velocity error and the true dispersion becomes (\citealt{Koposov_11}). Although we expect our velocity error to be an order of magnitude smaller than the true velocity dispersion in Fornax, we test the accuracy of our error estimates from stars with multiple, individual measurements.
For 15 stars in our sample we have 12 individual measurements respectively, and Figure\,\ref{fxcor_err_fig} compares the standard deviation from individual repeated measurements ($\sigma_{\mathrm{true}}$ to the mean error determined by \emph{fxcor} ($\sigma_{\mathrm{fxcor}}$) as a function of [Fe/H]. We find good agreement between these two numbers, with a mean ratio $\sigma_{\mathrm{fxcor}}/ \sigma_{\mathrm{true}}= 0.97\pm0.10$, and no trend with metallicity. 

\begin{figure}[htb]
\begin{center}
\includegraphics[width=0.5\textwidth]{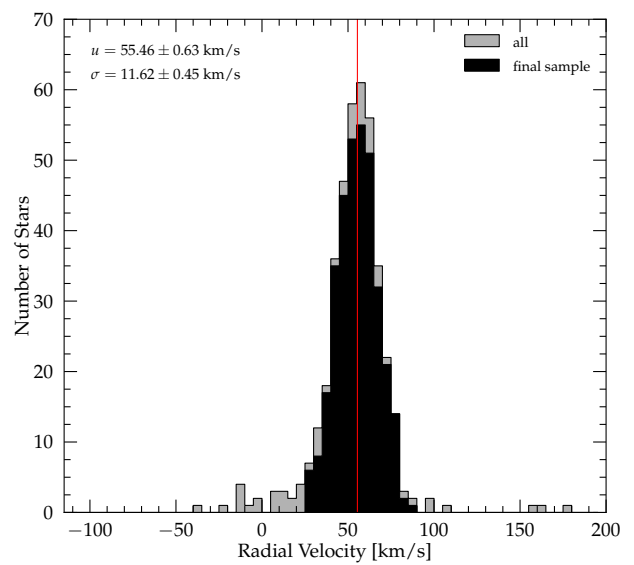}
\caption{Radial velocity distribution of stars in our sample. The red vertical line indicates the mean systemic velocity ($RV_{sys}=55.46$\,$\mathrm{km\,s^{\rm -1}}$) and stars that pass the iterative clipping procedure and the visual inspection described in the text are highlighted in black. The maximum velocity deviation of stars from the mean in order to have a likelihood of at least 99\% to be a Fornax member is $\pm30.70\,\mathrm{kms^{-1}}$. }
\label{fig_rv}
\end{center}
\end{figure}

The derived RVs can be used to assess the membership of each target star to Fornax and to weed out foreground stars. Previous studies have shown indications of an intrinsic velocity distribution in Fornax that deviates significantly from a Gaussian distribution (\citealt{Battaglia_06}). For this reason we make use of the biweight estimator (\citealt{Beers_90}, see also \citealt{Walker_06}), which is more robust against underlying non-Gaussian populations than a simple n-$\sigma$-clipping. However, its characteristic distribution width ($S_{BI}$) corresponds to a Gaussian standard deviation if the data are normally distributed. To reach a membership likelihood of 99\%, we clip the data at $2.58\times S_{BI}$, where $S_{BI}$ is redetermined in an iterative process until convergence.
See Figure \ref{fig_rv} for the distribution of RVs in our sample and a visualization of the clipping limits.

Next, we visually inspect our spectra and exclude those from the sample with either an apparent non-stellar origin (e.g., background galaxies, quasars, etc.) or spectra with strong telluric remnants within the environment of the three CaT lines.
Additionally, we only keep stars in our final sample with a minimum S/N per pixel of $\geq 10$, to guarantee reliable and accurate determination of velocities and CaT equivalent widths (EWs). 

Our target fields also cover two of the five known GCs (H2 and H5; \citealt{Hodge_61}) associated with Fornax. Because the chemical enrichment history of GCs can be significantly different from that of the field star population, we flag possible GC stars (those within $60\arcsec$ around the cluster centers) in our sample and exclude them in our chemical and dynamical analysis. See Section\,\ref{GCs} for a separate analysis of these stars and derived properties for the GCs.

Applying all selection criteria discussed here, our sample of bona-fide Fornax field stars consist of 378 stars, plus 13 possible GC members.

Located at a Galactic latitude of $b=-65.7\degree$ (\citealt{McConnachie_12}), we expect the foreground contamination for Fornax to be minimal (see also \citealt{Battaglia_06}). To estimate the number of foreground stars in our sample, we use the Besan\c{c}on Model for stellar population synthesis of the MW (\citealt{Robin_03}) and extract all synthetic field stars up to the distance of Fornax ($d=147\,\textrm{kpc}$) in a solid angle equivalent to our combined pointing area ($A=0.139\,\mathrm{deg^2}$) and within the same photometric selection box that we used for the initial target selection. We find $\sim30$ foreground stars matching these criteria. When we further consider the fraction of stars inside this box that were finally selected for spectroscopy, and furthermore take into account that the velocity clipping already rejects all stars with radial velocities outside of the clipping range, we expect only about a handful of foreground stars in our final sample, which is negligible for the further analysis.

To determine the systemic RV ($RV_{sys}$) and its intrinsic velocity dispersion ($\sigma_{sys}$) from the cleaned sample of Fornax field stars, we use the maximum-likelihood statistics described in \citet{Walker_06} which yield $RV_{sys}=55.46\pm0.64$\,$\mathrm{km\,s^{\rm -1}}$ and $\sigma_{sys}=11.84\pm0.45$\,$\mathrm{km\,s^{\rm -1}}$. 
These numbers are in good agreement with previous measurements from \citet{Battaglia_06} who found $RV_{sys}=54.1\pm0.5$\,$\mathrm{km\,s^{\rm -1}}$ and $\sigma_{sys}=11.4\pm0.4$\,$\mathrm{km\,s^{\rm -1}}$, or \citet{Walker_09} who obtained $RV_{sys}=55.2\pm0.1$\,$\mathrm{km\,s^{\rm -1}}$ and $\sigma_{sys}=11.7\pm0.9$\,$\mathrm{km\,s^{\rm -1}}$ from more evenly distributed sample within the tidal radius of the galaxy.
The mean velocity for both the southwestern and the northeastern field is practically the same within the uncertainties ($RV_{W}=54.50\pm 0.61$\,$\mathrm{km\,s^{\rm -1}}$ and $RV_{E}=56.26\pm 0.68$\,$\mathrm{km\,s^{\rm -1}}$, respectively), which supports previous findings that Fornax' rotational component is dynamically insignificant (\citealt{Walker_06}).

\subsection{CaT-metallicities}

The CaT is one of the most prominent absorption features in the near-infrared part of stellar spectra. Its three lines are located at $8498.02$\,\AA, $8542.09$\,\AA, and $8662.14$\,\AA, respectively.
Because the CaT line-strength varies as a function of metallicity it has been used as an indicator for $\mathrm{[Fe/H]}$ in a variety of galactic and especially extragalactic systems for which detailed high-resolution spectra in combination with high S/N are extremely time expensive. For our spectra with $R\sim16,000$ and a S/N typically around 30, the CaT can be used to derive metallicities for the large majority of stars. 

CaT metallicities are typically determined in two steps. At first the EWs of the three absorption features are derived by fitting the line profiles in a continuum-normalized spectrum with some analytic function. The next step is to transform the CaT EWs into [Fe/H]. Those calibrations not only relate the change in the CaT line profile as a function of iron abundance, but also remove effects of stellar atmospheric parameters, in particular $log\,g$.
Published calibration equations, which correlate the CaT EWs to the intrinsic $\mathrm{[Fe/H]}$ of a star, depend on the exact approach of measuring the CaT-EWs (see Section\,\ref{cat_iron} for a detailed discussion). 
In the past there have been different approaches to derive a star's $\mathrm{[Fe/H]}$ from the CaT. While originally all three lines have been used with equal weights (\citealt{Armandroff_88}, \citealt{Cole_04}), \citet{Rutledge_97a} apply a weighted sum of the lines to account for their different line strength. In recent years, however, most analyses are solely based on the two strongest ($CaT_{2}$, $CaT_{3}$) lines (e.g. \citealt{Koch_06}, \citealt{Battaglia_08}, \citealt{Starkenburg_10}) due to the apprehension that the weakest line adds more noise to the final result. We will follow that argumentation and restrict our analysis to $CaT_{2}$ and $CaT_{3}$.

Here, we adopt the respective line- and continuum bandpasses given in \citet{Armandroff_88} and also follow their approach in correcting for traces of a continuum trend in the vicinity of the lines by fitting a linear function through the median of each continuum bandpass to both sides of the line.

\begin{figure}[htb]
\begin{center}
\includegraphics[width=0.5\textwidth]{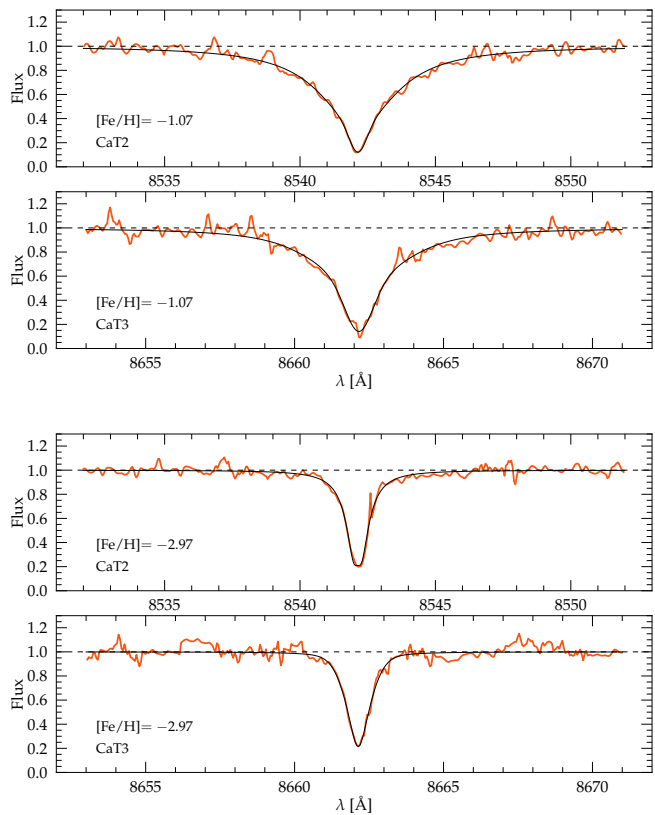}
\caption{Determination of the EW from the CaT, from continuum-corrected spectra in case of a metal-rich target (upper two panels) and a metal-poor star (lower two panels).
A combination of a Gaussian and a Lorentzian function (black line) is fitted to the absorption features in order to account for the strong damping wings, particularly for the metal-rich stars. The plotted wavelength range represents the adopted line bandpasses. The S/N of both targets is $\sim30$, a typical value for our sample.}
\label{fig_cat_fit_1}
\end{center}
\end{figure}

Next, we determine the EW from the actual line profile.
Usually, a simple Gaussian is not sufficient to model the shape of the lines appropriately, because it is significantly underestimating the broad damping wings of the CaT. This is specifically significant for strong lines and hence for high metallicities, which would consequently introduce an unwanted bias (\citealt{Rutledge_97a}).
On the other hand, a simple integration of the flux over the line bandpass (as originally performed by \citealt{Armandroff_88}) also sums weaker lines of other elements inside the interval which might show different dependencies on the metallicity and atmospheric parameters than the CaT.
While \citet{Rutledge_97a} use a Moffat function to account for the damping wings, \citet{Battaglia_08} and later \citet{Starkenburg_10} use a Gaussian with an additional empirical correction term defined by the integrated flux within the line. 
In our work, we use the sum of a Gaussian and a Lorentzian function to fit the CaT lines and determine the EW from numerical integration, which provides a good fit for both metal-poor and metal-rich stars (see Figure\,\ref{fig_cat_fit_1}). This approach has been used in several previous studies (e.g., \citealt{Cole_04}, \citealt{Koch_06}) as well as for the determination of CaT-[Fe/H] calibration relations (\citealt{Carrera_13}).

To ensure reliable results we visually inspect each fit and exclude stars where the function fails to reproduce a reasonable continuum level or the shape of the fit does not agree with physical expectations.
We can derive reliable EWs for 346 field stars and 13 additional stars which are likely GC members.

Finally, we use the recently published calibration equations of \citet{Carrera_13} to obtain [Fe/H] from the derived CaT-EWs. These authors made a dedicated effort to extend the classical calibration range of GC-based calibration from $\mathrm{[Fe/H]} \approx -2.0$ to metallicities as low as $-4.0$\,dex.

To obtain uncertainties for our CaT-metallicities we first use the covariance matrix of each line fit to determine $\sigma \mathrm{EW}$ from the individual uncertainties in each free fitting parameter and their dependencies. To propagate the error through the calibration equation we use the uncertainties for the individual calibration indices from \citet{Carrera_13} and estimate the uncertainty on the luminosity-normalization as $\sigma(V-V_{HB})=0.10$. From this, we find a median error for our CaT-metallicities of $\sigma \textrm{[Fe/H]}= 0.10\,\textrm{dex}$.
Although we find that the minimum uncertainty increases with metallicity due to the larger wings of the lines at high [Fe/H], which add more noise than signal, the mean uncertainty at each metallicity is about constant because the scatter towards higher $\sigma \mathrm{[Fe/H]}$ is larger for metal-poor stars .

As a crosscheck for our uncertainties derived from individual line fits, we make a second approach with the analytical formula proposed in \citet{Cayrel_88}, based solely on the S/N of the spectra as well as their spectral resolution: 

\begin{equation}
\sigma~\textrm{EW} = 1.725 \sqrt{\sigma_{Gauss}} \ {S/N}.
\label{eq1}
\end{equation}

\begin{figure}[htb]
\begin{center}
\includegraphics[width=0.5\textwidth]{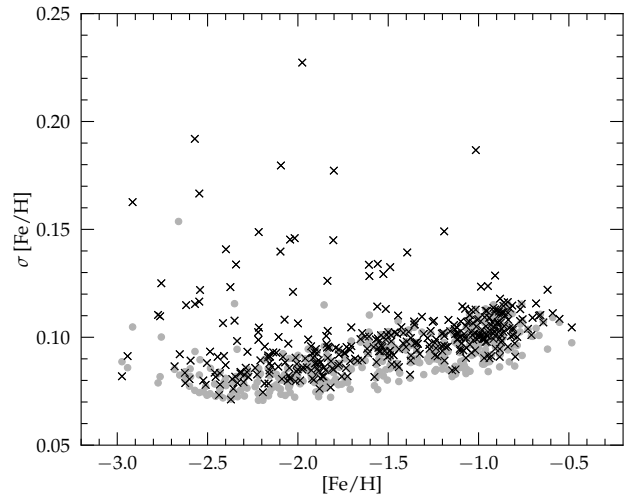}
\caption{Statistical uncertainty on our CaT metallicity values. Black crosses are errors obtained from MC-simulations on our line fit, gray points represent errors derived with the analytic formula from \citet{Cayrel_88}.}
\label{fig_error_cat}
\end{center}
\end{figure}

In the above equation, $\sigma \mathrm{EW}$ depends on the Gaussian width of the lines. Because a combined function of a Gaussian and a Lorentzian does not provide this information directly, we determine $\sigma_{Gauss}$ by fitting a \emph{pure} Gaussian to the absorption features. 
Both error estimates are shown in Figure \ref{fig_error_cat}, which shows that the analytic estimates are in good agreement to the uncertainties derived from the line-fitting covariance matrix. 

\subsection{The Mg\,I line at 8806.8\,\AA~as dwarf-giant index}

Recently, the neutral Mg line at 8806.8\,\AA~has been proposed as an indicator for stellar surface gravities, and hence to separate possible foreground dwarfs from RGB galaxy members (\citealt{Battaglia_12}). Since our spectra cover both the CaT and the Mg\,I-feature, we can identify additional foreground contamination, which has not been removed through the radial velocity clipping. We derive the Mg\,I EW by simple integration over a 3\,\AA-interval around the line center in the continuum-normalized spectra. \citet{Battaglia_12} use a broader (6\,\AA) interval around the line, but we find that this includes a contaminating Fe\,I line located at $8804.6$\,\AA, which we intend to avoid with the smaller integration corridor. Note that, when we apply the narrow interval, we do not cut off the wings even from the broadest Mg\,I-lines which do not span more than $\sim1.5$\,\AA~for stars in our sample.

\begin{figure}[htb]
\begin{center}
\includegraphics[width=0.5\textwidth]{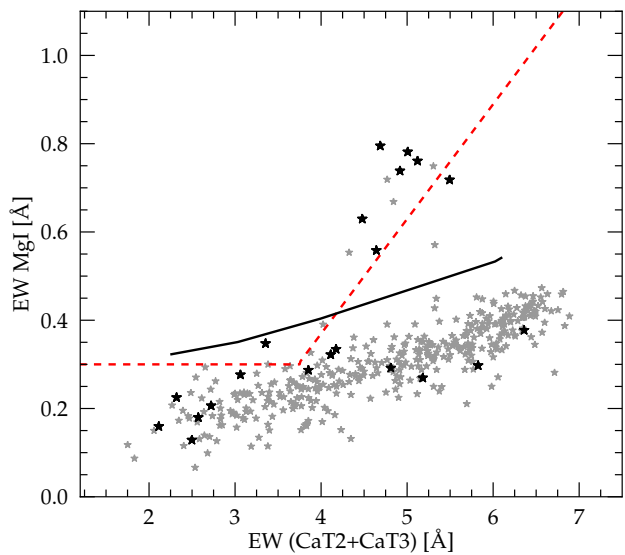}
\caption{EW of the two strongest CaT lines in comparison to Mg\,I at 8806.8\AA. Foreground dwarfs have higher log\,g and display a stronger Mg-line at given $\mathrm{[Fe/H]}$ compared to giant stars. Results from our sample are shown as gray symbols, and stars which previously have been excluded based on their velocities are highlighted in black. The data shows a clear sequence of RGB stars with a few (dwarf) foreground contaminants well above the general trend. The separation-criterium from \citet{Battaglia_12} is indicated as a red dashed line, but does not provide a good cut for our sample. Our own separation limit is indicated as a black solid line and removes only outliers $\geq0.15\AA$ above the median Mg linestrength at a given CaT EW.}
\label{fig_mag_indicator}
\end{center}
\end{figure}

In Figure\,\ref{fig_mag_indicator} the MgI EWs are plotted against the EWs from the two strongest CaT lines. We observe a group of obvious outliers with Mg\,I EWs more than 0.2\,\AA~above the majority of stars that are located on a well defined sequence, which indicates that the proposed method is generally a useful dwarf-giant separator. However, as can be seen in Figure\,\ref{fig_mag_indicator}, the separation function given in \citet{Battaglia_12} does not yield an optimal cut to these outliers, which may be explained by the different CaT EW fitting technique applied to their data (see Section\,\ref{cat_iron}).
We therefore decide to simply remove stars $\geq0.15\AA$ above the median Mg-EW at any given CaT-EW, which concerns five stars from our previous sample, in addition to the already flagged RV outliers. Note, that this number is in excellent agreement with our estimate based on comparison with the Besan\c{c}on Model (see Section\,\ref{radial_velocities}).

\subsection{The GCs H2 and H5}
\label{GCs}
The chemical composition, age, and dynamical information of extragalactic GCs give important clues about their evolution and the evolution of their host galaxy (\citealt{Brodie_06}). While their ages help to understand GC formation mechanisms (\citealt{van_den_Bergh_81}), detailed abundance analysis of their stellar content helps to constrain the chemical enrichment processes within the cluster and in the environment of their formation (see, e.g., \citealt{Gratton_04}). 

The GCs in Fornax have been studied intensively in the past and their metallicities and RVs have been determined with various methods. However, due to their limited spatial extent, most spectroscopic studies relied on integrated light analysis (\citealt{Strader_03}, \citealt{Larsen_12b}), and so far only \citet{Letarte_06} carried out a detailed abundance analysis for a small number of \textit{individual} stars in three of the clusters. 
Two Fornax GCs (H2 and H5) were included in our target fields and it is therefore likely that some stars in our sample are GC members. 
To identify bona-fide GC members, we first select stars within $60\arcsec$ (equivalent to the tidal radius) around the respective cluster centers: $(\alpha,\delta)=(02^{h}38^{m}40^{s}.1,-34\degree48\arcmin05\arcsec.0)$ for H2 and $(02^{h}42^{m}21^{s}.15,-34\degree06\arcmin04\arcsec.7)$ for H5. For all stars in question, we have reliable $\mathrm{[Fe/H]}$ abundances from the CaT as well as precise RV measurements. Unfortunately, the S/N for these stars is not sufficient to determine $\mathrm{\alpha}$-elements.

Besides a visual clustering of stars around the coordinates we find a striking similarity in metallicity and radial velocities for both sub-groups around H2 and H5, respectively (see Table\,\ref{table_GC}). When we exclude the star (ID 278) with significantly lower RV compared to the other candidates
\footnote{Note that we also find a significantly lower age for this star compared to the other candidates, which gives further support that it is not an actual member of H5 (see Section\,\ref{age_chapter})}
, we find $\mathrm{[Fe/H]}=-2.08\pm0.05$ and $RV=59.36\pm0.31$\,$\mathrm{km\,s^{\rm -1}}$ for H2 and $\mathrm{[Fe/H]}=-2.03\pm0.08$ and $RV=59.39\pm0.44$\,$\mathrm{km\,s^{\rm -1}}$ for H5. From our limited sample, these two systems have an identical metallicity and systemic RV within the uncertainties. 
Our numbers are in excellent agreement with previous findings: \citet{Larsen_12b} measures a metallicity of $-2.1\pm0.1$ and a radial velocity of $60.6\pm0.2$\,$\mathrm{km\,s^{\rm -1}}$ for H5 from integrated light spectroscopy and \citet{Letarte_06} obtained $\mathrm{[Fe/H]}=-2.10\pm0.1$ and $RV=63.8$\,$\mathrm{km\,s^{\rm -1}}$ from three individual stars in H2.
Note that for H2 we provide the largest sample of individual spectroscopic RV and [Fe/H] measurements today.



\begin{table}[htb]
\caption{Chemodynamical parameters for candidate GC members in our sample}
\label{table_GC}    
\centering
\begin{tabular}{lccccc}
\hline
\hline

			&					&				&	&		RV 		&	$\sigma$RV 	\\
\raisebox{1.5ex}[-1.5ex]{ID} & \raisebox{1.5ex}[-1.5ex]{GC} & \raisebox{1.5ex}[-1.5ex]{$\mathrm{[Fe/H]_{CaT}}$} & \raisebox{1.5ex}[-1.5ex]{$\sigma$[Fe/H]} &  [$\mathrm{km\,s^{\rm -1}}$]	 & [$\mathrm{km\,s^{\rm-1}}$] \\\hline

\phantom{1}94				&		H2					&		$-2.21$		&	0.11			&		67.18		&	0.94				\\
\phantom{1}95				&		H2					&		$-2.28$		&	0.09			&		61.91		&	1.29				\\
\phantom{1}97				&		H2					&		$-2.05$		&	$-$			&		62.31		&	0.76				\\
\phantom{1}99				&		H2					&		$-2.15$		&	0.09			&		60.04		&	1.43				\\
199				&		H2					&		$-1.95$		&	0.09			&		60.02		&	0.97				\\
201				&		H2					&		$-2.06$		&	0.08			&		62.98		&	1.06				\\
202				&		H2					&		$-1.92$		&	0.08			&		53.00		&	0.69				\\
203				&		H2					&		$-2.16$		&	0.09			&		56.75		&	0.87				\\
206				&		H2					&		$-1.91$		&	0.08			&		56.84		&	0.96				\\\hline
278 				&		None\tablefootmark{a}				&		$-1.87$		&	0.09			&		37.54		&	1.57				\\
423				&		H5					&		$-2.15$		&	0.09			&		58.93		&	0.76				\\
426				&		H5					&		$-2.05$		&	0.08			&		59.81		&	1.07				\\
427				&		H5					&		$-1.88$		&	0.08			&		59.43		&	1.52				\\\hline

\end{tabular}
\tablefoot{
\tablefoottext{a}{We have excluded ID 278 as a possible member for H5 due to its low line-of-sight velocity.} All other parameters for these stars are listed in Table\,\ref{table_field_stars_1}, \ref{table_field_stars_2}, and \ref{table_field_stars_3}.}
\end{table}

\section{Testing the CaT-calibration}
\label{cat_iron}

Originally, the CaT has been calibrated to GCs with known metallicity (e.g. \citealt{Armandroff_88}, \citealt{Rutledge_97a}, and more recently \citealt{Battaglia_08}, \citealt{Koch_06}, and \citealt{Carretta_09}).
The calibration limits -- set by the metallicity range of the GCs that were used -- then have been extended with open clusters towards higher [Fe/H] (\citealt{Cole_04}).
At this point, a linear relation between the strength of the CaT lines was assumed with a zero-point that is linearly correlated with the stellar luminosity and thus gravity.
Recently, extensive tests have shown that both correlations show non-linear trends when large ranges of either [Fe/H] and/or luminosity are sampled 
(\citealt{Battaglia_08}, \citealt{Starkenburg_10}).
In the last years, \citet{Starkenburg_10} and \citet{Carrera_13} developed new CaT-calibrations, which both add quadratic terms to the equations with the goal to extend the acceptable calibration range to both sides in [Fe/H], and particularly towards more metal-poor stars in order to remove the existing bias from the metal-poor tail in extragalactic metallicity distribution functions.

We have a large, homogeneous sample of stars with sufficient spectral resolution and S/N in order to determine $\mathrm{[Fe/H]}$ \emph{independently} from both the CaT and from detailed analysis of individual Fe-lines in our spectra. This provides a unique opportunity to test the different existing CaT-calibrations over a range of more than 2\,dex from $\mathrm{[Fe/H]}=-2.8$ to $-0.5$.
In the following, and in the remainder of this work we will refer to $\mathrm{[Fe/H]}$ measured from the CaT as \emph{CaT-metallicities}, while \emph{Fe-metallicities} indicate iron abundances derived from individual iron lines. For a detailed description of the latter, see \citet{Hendricks_14}

Here, we test three different equations to calibrate our EWs to $\mathrm{[Fe/H]}$. 

\begin{itemize}

\item i) A classical GC-calibration from \citealt{Koch_06}:

\begin{equation}
\mathrm{[Fe/H]}=-2.77+0.38 W',
\end{equation}
with
\begin{equation}
W'=EW_{2+3} + 0.55 (V_{\star}-V_{HB}),
\end{equation}
where $EW_{2+3}$ denotes the sum of the two strongest CaT lines, and $(V_{\star}-V_{HB})$ is the relative V-band magnitude of a star above the horizontal branch.

\item ii) The semi-synthetic calibration from \citet{Starkenburg_10}, who take into account the non-linear behaviour of the EWs by adding a quadratic term to the calibration equation which is derived from synthetic line analysis:

\begin{eqnarray}
\mathrm{[Fe/H]}&=& a + b \times (V_{\star}-V_{HB}) + c \times \sum EW_{2+3}   \\ 
&+& d \times (\sum EW_{2+3})^{-1.5} \nonumber \\
&+& e \times  \sum EW_{2+3} \times (V_{\star}-V_{HB}) \nonumber
\label{eq4}
\end{eqnarray}
\label{eq4}

with $$(a,b,c,d,e) = (-2.87, 0.195, 0.458, -0.913, 0.0155)$$.

\item iii) The most recent calibration from \citealt{Carrera_13}, who use a combination of open clusters, GCs, and metal-poor field stars to derive a purely empirical calibration following the same non-linear form as given in Eq.\,4. They find $(a,b,c,d,e) = (-3.45, 0.11, 0.44, -0.65, 0.03)$. According to the relative linestrength of the three CaT features given in \citealt{Carrera_13}, we have to divide each EW-term by $0.81$ in order to account for the fact that we only use the two stronger CaT-lines, and not all three as it is done in their paper. 

\end{itemize}

Note that in i) and iii), CaT EWs are fitted with a sum of a Gaussian and Lorentzian, which is also our approach to measure the linestrength, whereas ii) uses an analytic correction term to an initial Gaussian fit as described in \citet{Battaglia_08}.

\begin{figure}[htb]
\begin{center}
\includegraphics[width=0.5\textwidth]{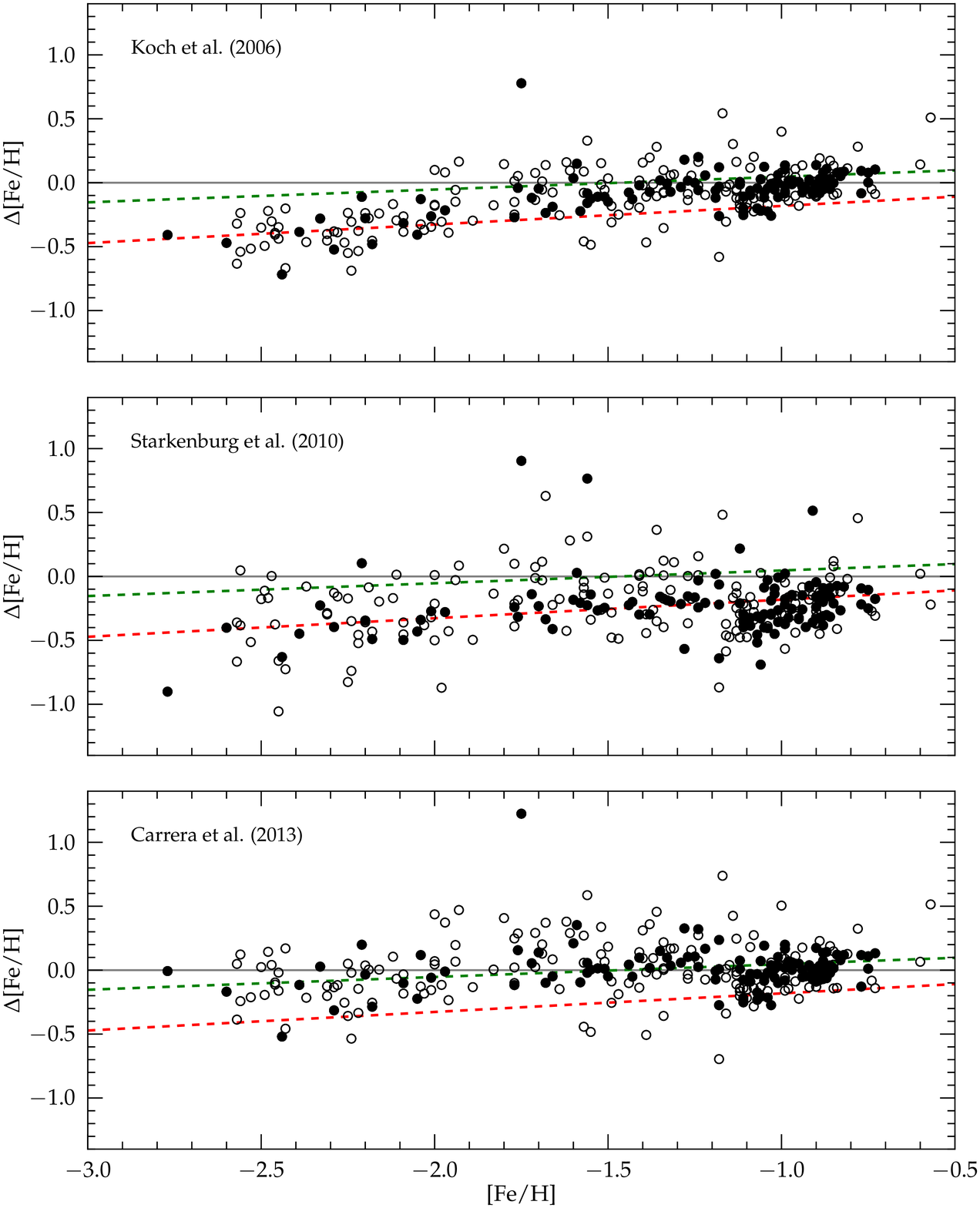}
\caption{Comparison of different CaT-[Fe/H] calibrations to independent Fe-abundances from iron-line analysis for the same sample of stars. In each panel $\Delta \mathrm{[Fe/H]}= Fe-CaT$. We show the results for our complete sample in open circles and highlight stars with $S/N\geq25$ and $\sigma\mathrm{[Fe/H]}\leq0.15$ with filled symbols. The red and green dashed line in each panel indicate the best linear fit to the Starkenburg- and Carrera-calibration, respectively, and reveals a constant $\sim 0.2$\,dex-offset between them, where the latter one yield the better fit to our data. The classical GC-calibration in the top panel shows a clear systematic trend for $\mathrm{[Fe/H]}\leq-1.8$, resulting in CaT-metallicities too metal-rich by as much as $0.5$\,dex.}
\label{fig_CaT_calibration}
\end{center}
\end{figure}

The results from the different calibrations are shown in Figure\,\ref{fig_CaT_calibration}.
 We find, in agreement with \citet{Battaglia_08}, that a classical GC-calibration is only valid between $-1.8\leq\mathrm{[Fe/H]}\leq-0.6$, and shows strong deviations at lower metallicities, where the CaT-metallicity is becoming systematically too metal-rich by $\geq0.5$\,dex (see also \citealt{Koch_08}).
When we use the calibration of \citet{Starkenburg_10}, we do not observe a significant trend in our CaT-metallicities compared to the Fe-metallicities at low $\mathrm{[Fe/H]}$. However, there is a zero-point offset between the two approaches. When we fit a linear function to the residuals ($\Delta \mathrm{[Fe/H]}=x_{1}\times \mathrm{[Fe/H]_{HR}}+x_{0}$), we obtain $x_{1}=0.15\pm0.05$ and $x_{0}=-0.04\pm0.07$ as best fitting parameters, indicating a negligible dependance on $\mathrm{[Fe/H]}$, but with an offset of $\sim0.2$\,dex at $\mathrm{[Fe/H]}= -1.5$, resulting in too metal-rich CaT-metallicities. 
Finally, the Carrera-calibration equations agree remarkably well with our Fe-metallicities. As can be seen in the lowest panel in Figure\,\ref{fig_CaT_calibration}, there is neither a dependance of the derived CaT-metallicity on $\mathrm{[Fe/H]}$, nor a significant zero-point shift as observed for the Starkenburg-calibration. A similar fit of a linear function gives best fitting parameters of $x_{1}=0.10\pm0.04$ and $x_{0}=0.15\pm0.06$, corresponding to an exact match at $\mathrm{[Fe/H]}=-1.50$\,dex.

\begin{figure}[htb]
\begin{center}
\includegraphics[width=0.5\textwidth]{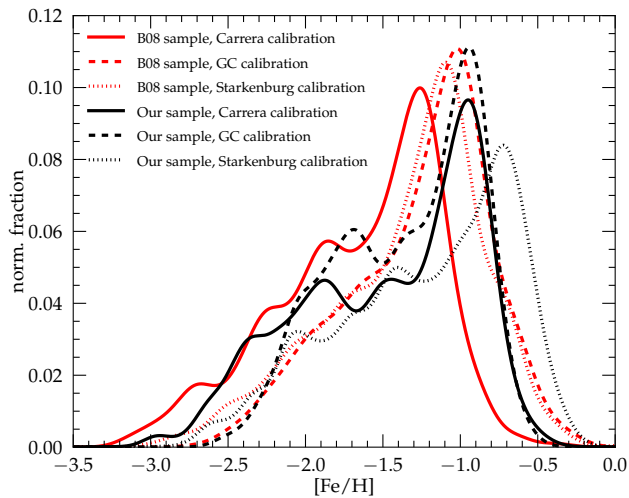}
\caption{Comparison between different CaT-[Fe/H] calibrations applied to two independent datasets. The B08-sample has been analyzed with an empirical correction to a Gaussian fit, whereas our EWs have been determined from a combined Gaussian and Lorentzian function. Analyzed from the position of the prominent peak in the distribution, a good agreement for the zero-point in [Fe/H] between both sets is obtained when the line fitting technique in the calibration relation and the scientific sample is the same. In contrast, different approaches in the line fitting technique can result in significant offsets of more than 0.5\,dex between the different samples.}
\label{calibration_offsets}
\end{center}
\end{figure}

To investigate the origin of the zero-point offset between the two most recent calibration equations, we make use of the CaT-catalog from \citet{Battaglia_08} -- an extended version of the catalog published in \citet{Battaglia_06} -- for which these authors provided not only [Fe/H], but also the underlying EWs (from here on ``B08-sample'', G. Battaglia, priv. comm.).
When we compare the MDF derived from the different datasets we find excellent agreement for the dominant and narrow peak metallicity in the MDF at $\sim-1.0$\,dex (see also Section\,\ref{mdf_chapter}) when we use the Carrera-calibration or the \citet{Koch_06} GC-calibration for our data, and the Starkenburg-calibration or the \citet{Battaglia_08} GC-calibration for the B08-sample. In contrast, when we apply the Starkenburg-scale to our data, the peak appears $\sim 0.2$\,dex too metal-rich. Vice versa, the peak in the B08-sample becomes too metal poor by the same amount when we apply the Carrera-calibration on their EWs (see Figure\,\ref{calibration_offsets}). 
Strikingly, both calibrations for which our sample peaks at $\sim-1.0$\,dex use the sum of a Gaussian and a Lorentzian to fit the line profiles, as we do to derive our EWs. Similarly, the calibrations that bring the B08-sample to peak at the same metallicity applied an empirical correction to a Gaussian fit, corresponding to the approach in \citet{Battaglia_08}.
We therefore conclude that the actual fitting approach for the CaT absorption features can have significant effects on the derived EWs and is most likely the reason for the 0.2\,dex-offset between the Carrera- and the Starkenburg-calibration equations, when applied to our sample.

In other words, we find that both the Starkenburg- and the Carrera-calibration show good agreement with high-resolution results between $-3.0\leq \mathrm{[Fe/H]}\leq -0.5$, but only if EWs are derived with the corresponding method to the applied calibration. Otherwise, systematic offsets in the order of $\sim0.2$\,dex in the derived CaT-metallicity can be introduced at all metallicities. This could result in systematic discrepancies of up to $\sim0.5$\,dex between independent CaT-studies. 

\section{The Age-Metallicity Relation}
\label{age_chapter}

An accurate age-metallicity relation is a powerful tool to determine a galaxy's chemical enrichment history (e.g., \citealt{Carraro_98}, \citealt{Haywood_13}). Unfortunately, the determination of stellar ages from isochrone fitting suffers from several systematic and statistical uncertainties
On the one hand, the position of a star in the CMD depends on both chemical composition and age, so that precise, individual $\mathrm{[Fe/H]}$ and $\mathrm{[\alpha/Fe]}$ information is required to break this degeneracy. On the other hand, assumptions for the distance modulus and interstellar reddening and extinction are necessary and therefore pose a significant source of systematic uncertainty. 
The analysis is additionally based on the assumption that a given set of isochrones correctly predicts stellar evolutionary sequences for stars of given age and chemistry and consequently relies -- among others -- on assumptions for mixing length, core-overshooting, and mass loss in the models.
On top of these sources of error, the age-sensitivity of stellar positions on the RGB is weak, and large random errors are introduced from even small uncertainties in the stellar photometry. This is specifically true for old populations where the uncertainties can exceed several Gyr.
Therefore we dedicate a separate discussion to the individual statistical and systematic uncertainties governing relative age estimates in Section\,\ref{section_age_uncertainties}.

Previous studies like those of \citet{Pont_04}, \citet{Battaglia_06}, \citet{Lemasle_12}, or more recently \citet{de_Boer_12} make use of individual metallicity-measurements for large samples of stars in dSphs to derive their AMR. Other studies like \citet{del_Pino_13} use only photometric data and try to break the age-metallicity degeneracy by finding the best model fit for a wide grid of possible age-[Fe/H] combinations. From our sample, we not only have detailed CaT- and Fe-metallicities for the majority of our stars, but also know $\mathrm{\alpha}$-element abundances for some of them which enables us to reconstruct the $\mathrm{\alpha}$-enrichment for stars at a given $\mathrm{[Fe/H]}$ (see \citealt{Hendricks_14}). Therefore we are -- for the first time -- able to derive stellar ages in Fornax from isochrones \emph{individually} tailored for the stellar $\mathrm{[Fe/H]}$ \emph{and} $\mathrm{[\alpha/Fe]}$.

Precise photometric information is required in order to obtain a reasonable statistical uncertainty on stellar ages when measured for RGB stars. Originally, the main purpose of our photometry was the target selection, and consequently the photometric precision is lower than in dedicated photometric studies. For stars in our sample typically $\sigma (V-I) \approx 0.06$ and $\sigma V \approx 0.05$, which is too poor for a detailed age analysis (see Section\,\ref{section_age_uncertainties}). Therefore we make use of the recently published photometric catalog from \citet{de_Boer_12}, which covers the entire field of Fornax and provides $B$ and $V$ magnitudes for stars from the tip of the RGB down to the MSTO at $V\approx23.5$. For stars in our magnitude range, their photometric precision is typically $\sigma (B-V) \approx 0.005$ and $\sigma V \approx 0.004$, and thus an order of magnitude better than our own photometric information. When we allow for a maximum astrometric deviation of $\delta (RA,Dec) = 1\arcsec$, we are able to match $\geq95$\% of all stars in our sample with a star in the de Boer catalog.

For our age-analysis, we use the Dartmouth-isochrone database\footnote{http://stellar.dartmouth.edu/models/index.html} (\citealt{Dotter_08}), which provides stellar evolutionary sequences for $-2.5\leq \mathrm{[Fe/H]} \leq +0.5$ over a wide range of $\mathrm{\alpha}$-abundances ( $-0.2\leq \mathrm{[\alpha/Fe]} \leq +0.8$). We use their $\mathrm{[Fe/H]}$-interpolation program to generate isochrones for the exact stellar CaT-metallicities and assign an $\mathrm{[\alpha/Fe]}$ which is closest to the value we determined for our spectra. Note, that the spacing in their grid of $\mathrm{\alpha}$ is $0.2$\,dex, so that we can anticipate a maximum discrepancy of $0.1$\,dex between the isochrone and the actual stellar value, which we assign by placing its CaT-metallicity on the empirical fiducial evolutionary $\mathrm{\alpha}$-sequence for Fornax determined in \citet{Hendricks_14}.

The foreground reddening in the direction to Fornax is low ($E(B-V)\approx0.03$). However, from the reddening maps provided in \citet{Schlegel_98} we find that there is some fluctuation within the field of Fornax with peak-to-peak differences as large has $\delta E(B-V)\approx0.05$ when the entire area within its tidal radius is assessed, and $\delta E(B-V)\approx0.015$, within the area of the two fields covered by our sample. Although these numbers appear small at first, it is important to note that they introduce a bias in the photometric color several times larger than the intrinsic photometric errors, and therefore can cause systematically different ages of $\geq1$\,Gyr for stars at different position in the galaxy, if a constant value is assumed for all of them (see Section\,\ref{section_age_uncertainties}).
To avoid such systematics, we use the Schlegel et al. reddening maps to determine an \emph{individual} reddening value for each star in our sample, based on its astrometric position. The V-band extinction is then computed assuming a standard reddening law, so that $A(V)_{\star}= 3.1\times E(B-V)_{\star}$.
Note that the resolution of the reddening maps ($\mathrm{FWHM}=6.1\arcmin$) provides information for $\sim13$ individual positions in each field. Individual reddening values have then been derived through numerical interpolation.

Finally, individual ages are determined through linear interpolation of the age-color relation at the corresponding V-band magnitude of the star, providing continuous results despite the discrete grid of isochrone ages. Since the isochrones only provide fiducial evolutionary tracks for $\mathrm{[Fe/H]}\geq-2.50$, but Fornax hosts a significant number of stars below that limit, we additionally derive lower age limits for stars between $-3.0\leq \mathrm{[Fe/H]}\leq-2.5$, by adopting the most metal-poor isochrone available for these stars. 
Here, we use a distance modulus of $(m-M)_{0}=20.84$, corresponding to a distance of $147$\,kpc, adopted from the most recent measurement of \citet{Pietrzynski_09}.

\begin{figure*}[htb]
\begin{center}
\includegraphics[width=1.0\textwidth]{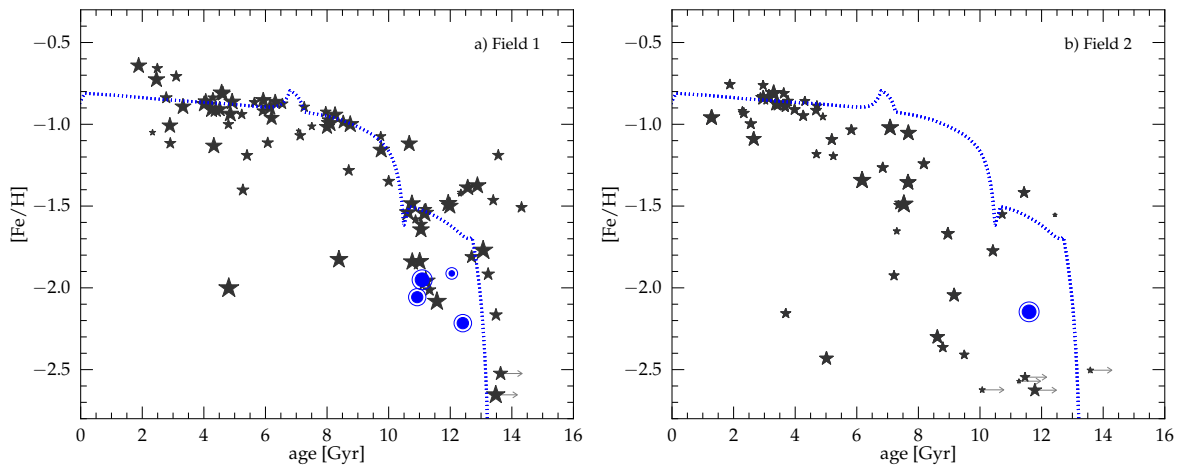}
\caption{Age-metallicity relation for Fornax field stars within the two distinct fields of our study. \textbf{Left:} South-Western field centred at $(\alpha,\delta)=(02^{h}38^{m}06^{s}.5, -34\degree49\arcmin52\arcsec.7)$, \textbf{Right:} North-Eastern field centred at $(\alpha,\delta)=(02^{h}41^{m}49^{s}.6, -34\degree03\arcmin55\arcsec.5)$. The observed differences between the AMR of Field\,1 and 2 are most likely due to small zero-point variations in the photometry (see text).
Symbol sizes in both panels reflect the statistical precision of the photometry and [Fe/H], where larger symbols indicate a better quality. Symbols with a right-handed arrow label additional stars with $-3.0 \leq \mathrm{[Fe/H]} \leq -2.5$, which fell outside the lower isochrone range and for which we adopted $\mathrm{[Fe/H]}=-2.5$ to derive lower age limits.
Blue symbols highlight stars which we identified as members of the GC H2 (left panel) and H5 (right panel), respectively.
The dotted blue line in both panels shows the model prediction for the best fitting SFH from \citet{Hendricks_14}, when we assume the age of Fornax to be $13.3$\,Gyr. While stars from Field\,2 appear systematically younger than the model, the AMR from Field\,1 shows excellent agreement with the predicted sequence. In contrast, in order to obtain a reasonable fit for Field\,2, Fornax could not be significantly older than $10$\,Gyr, a scenario which can be ruled out from previous photometric and spectroscopic age estimations.
From the well populated AMR of Field\,1, we find a fast enrichment in $\mathrm{[Fe/H]}$ until $\sim 8$\,Gyr ago, after which the enrichment becomes extremely shallow, so that a large range of ages accumulate at $\mathrm{[Fe/H]}\approx -1.0$. Finally the chemical enrichment seem to steepen again for stars younger than $4$\,Gyr, indicating a sudden increase in SF activity at this time.}
\label{fig_age_fe_relation}
\end{center}
\end{figure*}

In Figure\,\ref{fig_age_fe_relation}, we show the resulting AMR for Fornax. Because the age-precision fundamentally depends on the photometric quality and the uncertainty in [Fe/H], we only show stars for which $\sigma (B-V) \leq0.01$ and $\sigma \mathrm{[Fe/H]_{CaT}}\leq0.15$. We further restrict our sample to stars with $V\leq19.5$\,mag, due to a significantly lower age-sensitivity of isochrones at fainter magnitudes. Finally, we exclude all stars for which $\Delta ( \mathrm{[Fe/H]_{CaT}} - \mathrm{[Fe/H]_{Fe}}\geq0.3$. 

Overall, we find a clear correlation between age and [Fe/H], where older stars become more metal-depleted. 
The very few young, but metal-poor stars can most likely be assigned to a small fraction of either foreground stars or AGB interlopers which may be present in our sample.
It is striking that the detailed chemical enrichment appears different for the two distinct fields from which our sample is taken. The most likely origin of the systematic difference in individual stellar ages is a zero-point difference within the photometry, since the angular separation between Field\,1 and 2 is almost 1\degree, and the photometry for each field thus originates from different pointings. Therefore, small variations in the photometric zero-points for each field can be expected. Specifically, we find that a constant shift of only $\Delta (B-V)\approx0.05$, applied to one of the fields, is sufficient to bring the AMR of both fields in agreement (see Section\,\ref{section_age_uncertainties} for a detailed discussion).

For Field\,1, we find an extremely fast enrichment for older ages until about $\sim8$\,Gyr, after which the enrichment becomes shallow and $\mathrm{[Fe/H]}$ only increases by $\sim 0.3$\,dex over $\sim6$ Gyr, indicating a non-linear evolution of $\mathrm{[Fe/H]}$ with time. Finally the chemical enrichment seems to steepen again for stars younger than $4$\,Gyr, which is a hint for a sudden increase in SF activity at this epoch. Interestingly, this AMR shows similarities with the chemical enrichment pattern of the Magellanic Clouds (\citealt{Pagel_98}). However, the SFH of Fornax has to be somewhat different from these systems, given the larger fraction of old, metal-poor stars in Fornax compared to both the Large- and the Small Magellanic Clouds (\citealt{Cole_00}, \citealt{Dobbie_14}).
From Field\,2 the evolution of [Fe/H] appears slightly more linear with only a smooth flattening in chemical enrichment over time and generally younger ages.
As we will discuss below, a comparison to chemical evolution models favour the ages derived from Field\,1 to be the sample with the more accurate photometric calibration.
The observed scatter in the AMR of both fields can be caused by statistical uncertainties in the photometry and chemical abundances of the stars, or it can be introduced by systematic outliers like foreground dwarfs and AGB stars. Taking these effects together, the observed scatter does not give a hint for a significant spread in Fornax' AMR.

Chemical evolution models for dSphs as presented, e.g., in \citet{Lanfranchi_03} naturally predict the AMR for an assumed SFH of a galaxy.
In a recent study, these models have been used to constrain the SFH in Fornax from chemical element information (\citealt{Hendricks_14}).
In Figure\,\ref{fig_age_fe_relation}, we overplot the predictions for our best-fitting evolutionary scenario from that work. This model is characterized by three major SF bursts and an increasing SF efficiency over time. 
Generally the model predicts an extremely fast enrichment at early times, and an almost flat $\mathrm{[Fe/H]}$-plateau after the first few Gyr, as we observe in the data. When we set the age of Fornax in the model to $13.3$\,Gyr, we find an excellent agreement with the AMR as seen in Field\,1 while ages derived from stars in Field\,2 appear systematically younger than the model prediction. The age assumption for Field 1 seems reasonable when compared to previous studies that consistently found stars $\geq12$\,Gyr in Fornax (\citealt{Battaglia_06}, \citealt{de_Boer_12}).
In contrast, to obtain a reasonable model agreement for Field\,2, the assumed age of Fornax needs to be $\leq10$\,Gyr, a scenario which can be ruled out from previous photometric and spectroscopic age estimations of Fornax' oldest populations. Specifically, the lower age limit of Fornax should be constrained by its GC population, and at least three of the five globulars have ages between $12$ and $13$\,Gyr (\citealt{Strader_03}).

Here, we are also able to derive ages for one star in the GC H5 and four stars in H2 (blue symbols in Figure\,\ref{fig_age_fe_relation}).
As expected, all stars from both clusters have about the same age ($\sim 12$\,Gyr), showing that the statistical error in our age analysis is reasonably small, even for old, metal-poor stars.
The GCs fall on top of the observed sequence from the field stars, indicating a similar early chemical enrichment of the ISM out of which the clusters and the field population formed. Here, it would be clearly of interest to have precise [Fe/H] and age information for the GC H4, which is presumably significantly more metal-rich and possibly younger than the remainder of Fornax' GC population. Placed in the AMR of the field stars, this GC could help to answer the question whether the proto-GC material enriched in the same way as the field of the galaxy. Note, that star 278, that we dubbed as a field star in the line-of-sight to GC H5 from its RV signature, also has a significantly younger age estimate (7.40\,Gyr) in our analysis, which supports our previous assumption that it is not an actual member of H5.

Finally, it is worth noting that we do not observe any difference in the AMR when we split our sample into subgroups of stars with $r_{ell} \leq 0.6\degree$ and $r_{ell} \geq 0.6\degree$, and therefore do not see signs for a differential chemical enrichment at different galactocentric distances. However, we cannot rule out such differential effects due to our limited radial coverage and it would be clearly desirable to have a spectroscopic sample of stars with accurate, homogeneous photometry over the whole tidal range to test this hypothesis.

\subsection{Chemical vs. photometric SFH}
The star formation history of Fornax has been studied recently by \citet{de_Boer_12} and \citet{Weisz_14}, both using a photometric approach to derive the most likely scenario from synthetic CMD fitting. While \citet{de_Boer_12} use ground-based photometry with an extended spatial coverage from the center of the galaxy out to $\sim0.8\degree$, the results of \citet{Weisz_14} are based on specifically deep HST photometry of Fornax, that -- however -- only covers the central parts of the galaxy.
Both (photometric) studies find an extended SFH for Fornax, starting $\geq12$\,Gyr ago and lasting to very recent times, with a fairly constant SF rate during most of this period. Interestingly, \citet{de_Boer_12} additionally report a radial variation in the SFH within the galaxy in a way that in the outer parts a larger fraction of stars have been formed at early times, compared to the SFH in the central region.

While the above mentioned studies report a purely empirical SFH from photometry, the scenario proposed by us is based on a physical model adjusted to fit the chemical properties of the galaxy. The observed AMR in our study supports a SFH with extended episodes of SF interrupted by short periods of low activity and characterized by an increasing SF efficiency over time, as has been used in our previous paper to fit the chemical evolution of $\mathrm{\alpha}$-elements and the MDF in Fornax (see \citealt{Hendricks_14}). The AMR additionally shows evidence of an increase in SF activity $\sim4$\,Gyr ago, seen as a sudden increase in [Fe/H] thereafter.

\begin{figure}[htb]
\begin{center}
\includegraphics[width=0.5\textwidth]{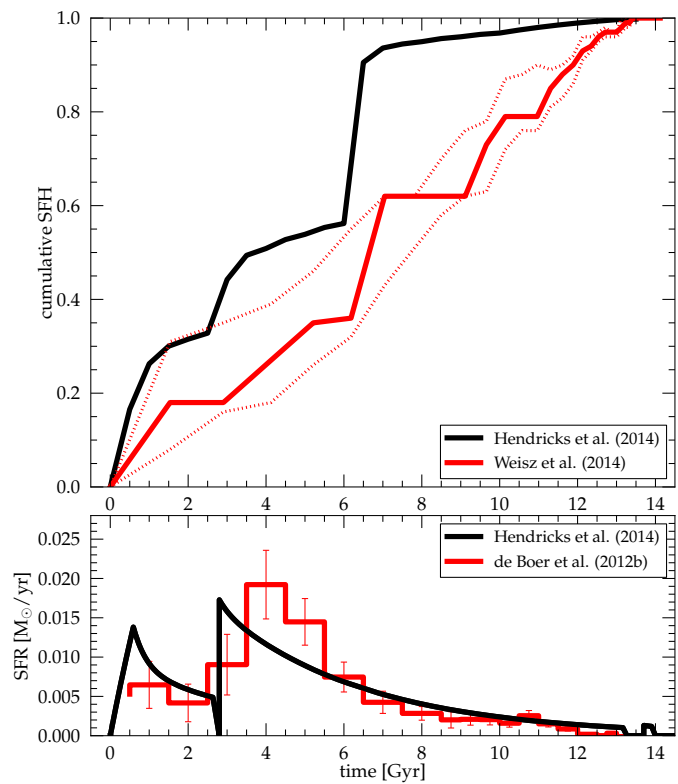}
\caption{\textbf{Top:} Comparison between the photometric SFH from \citet{Weisz_14} (red solid line) and our proposed evolutionary scenario (black solid line) based on the chemical evolution for the alpha-elements, the MDF and the observed AMR in the outer parts of Fornax. The red-dotted lines show the uncertainty interval from the photometric SFH. Note, that the larger fraction of stars formed at early times as predicted by our model can be expected from a radial SF gradient (see text for details).
\textbf{Bottom:} Comparison between the photometric SFH from \citet{de_Boer_12} and our model. The photometric history represents the results obtained from the outermost radii which overlaps with our sample, rescaled to match the total number of stellar mass formed in our scenario.}
\label{fig_new}
\end{center}
\end{figure}

Because the available gas which serves as star forming material within the galaxy decreases over time, our proposed scenario with an increasing SF efficiency predicts a roughly constant SF rate, as also observed in \citet{Weisz_14}. From Figure\,\ref{fig_new} it is clear that there is no fundamental difference between these two scenarios. 
However, our model predicts a generally larger fraction of stars produced at old times, a trend which can be explained, when taking into account the different radial positions of the samples and the radial trend in Fornax' SFH: while the photometric sample from \citet{Weisz_14} exclusively evaluates the central parts of the galaxy, our sample observes exclusively the outer parts, where a shift towards earlier SF is expected (\citealt{de_Boer_12}). 
Compared to the Weisz et al.\, SFH, our model also completes its SF $\sim5$\,Gyr earlier than observed in the photometric results. Part of this discrepancy may be explained with the above mentioned radial variations. In addition, such a lack of recent SF -- and consequently a lack of young stellar populations -- predicted by the model can be compensated with an additional episode of SF triggered by external effects like a merger event or the re-accretion of previously expelled gas. Such environmental impacts on the evolution of Fornax cannot be taken into account in our simple leaky-box model. Such a scenario would not only account for the discrepancies between the predicted and observed SFH, but simultaneously could explain the observed sudden increase in [Fe/H] at this time within the galaxy, whereas our model does not predict this evolutionary feature.
Our chemical evolutionary scenario is therefore in good agreement with the observed SFH in \citet{Weisz_14}, when it is supplemented with a late SF episode triggered by environmental interactions and when the radial gradient in SF within the galaxy is taken into account.

Finally, the time-resolved SFR shown in \citet{de_Boer_12} is in excellent agreement with our model, when we consider only the radial areas which overlap with the spatial extend of our sample (i.e., annuli 4 or 5 in their paper). Their observations, as well as our scenario, predict a slightly higher SFR at early times, until $\sim9.5$\,Gyr ago, after which the SFR drops continuously (see Figure\,4 in \citealt{Hendricks_14}).
Hereby it is important to stress that short gaps ($\leq1$\,Gyr) between the bursts of SF -- as proposed in our scenario -- are not in conflict with the continuous SFH derived from synthetic CMD fitting, because these studies are unlikely to resolve such short-time variations (\citealt{de_Boer_12a}).

\subsection{Signs for different dynamical populations}

Signs for dynamical peculiarities within the population of Fornax field stars have been reported in several previous studies. \citet{Battaglia_06} were the first who reported a larger velocity dispersion in the central region of the galaxy with bimodal RVs amongst metal-poor stars compared to the more enriched populations, and also compared to populations at larger radii. Later, a significant variation in the velocity dispersion between the metal-rich and the metal-poor stellar component has been confirmed by \citet{Walker_11}. Subsequently, \citet{Amorisco_12} found that, when stars with different metallicities are split into three subgroups, each of them show signs of a distinct dynamical behaviour, leading to the conclusion that a merger event in the past (preceded by a ``bound-pair'' scenario) may be a possible explanation.

\begin{figure}[htb]
\begin{center}
\includegraphics[width=0.5\textwidth]{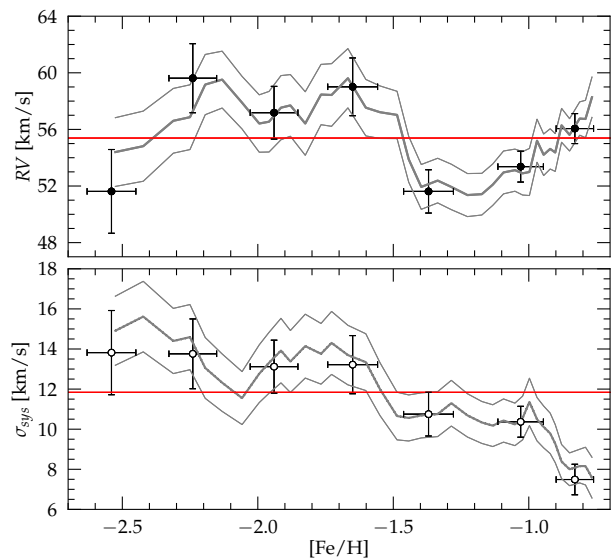}
\caption{\textbf{Top:} Radial velocities over the full range of $\mathrm{[Fe/H]}$ covered by our sample. The thick and thin solid lines represent the floating mean and the $1\sigma$-uncertainty interval for subsamples of 80 stars. Measurements for seven independent metallicity intervals with 0.3\,dex in size have been additionally taken and are shown as black dots. While the vertical error-bars indicate the intrinsic uncertainty for each bin, the horizontal bars represent the standard deviation of $\mathrm{[Fe/H]}$-values in each bin. The red line marks the mean systemic RV for the whole sample. We do not observe a significant trend in RV with $\mathrm{[Fe/H]}$, but the data show signs that there exist different velocity subgroups. \textbf{Bottom:} Same as in the top panel, for the velocity dispersion. A clear trend is visible with higher $\sigma_{sys}$ towards lower $\mathrm{[Fe/H]}$.}
\label{fig_velocity_dispersion}
\end{center}
\end{figure}

As can be seen in Figure\,\ref{fig_velocity_dispersion}, we also observe complex dynamics in the outskirts of Fornax.
Using the CaT-metallicities, we determine both the mean RV and the intrinsic velocity dispersion as a function of $\mathrm{[Fe/H]}$, following the same algorithm described in Section\,\ref{radial_velocities}. We find, that the velocity dispersion steadily increases from $\sigma_{sys}\approx7.5\,\mathrm{kms^{-1}}$ for $\mathrm{[Fe/H]}\geq-1.0$, to dispersions as high as $15\,\mathrm{kms^{-1}}$ for the most metal-poor stars. 
While different velocity dispersions for stellar populations in Fornax have been previously reported, here we are able to show that such variations do not only concern a specific population in the galaxy, but rather be part of a continuous trend from the most metal-poor to the most metal-rich stellar components in Fornax.
Such a dynamical pattern would be expected for a tracer population in a dark matter dominated halo, undergoing an outside-in SF.

Figure\,\ref{fig_velocity_dispersion} also indicates that individual metallicity subgroups have significantly different systemic line-of-sight velocities. To stress this fact, we overplot seven distinct subsamples and their intrinsic uncertainties to the floating mean, which indicates that stars between $-2.3 \leq \mathrm{[Fe/H]}\leq -1.5$ display a larger RV than the rest. However, because our sample is locally constrained, it is possible that we do not observe global trends with [Fe/H], but instead local inhomogeneities within the galaxy.

\begin{figure*}[htb]
\begin{center}
\includegraphics[width=1.0\textwidth]{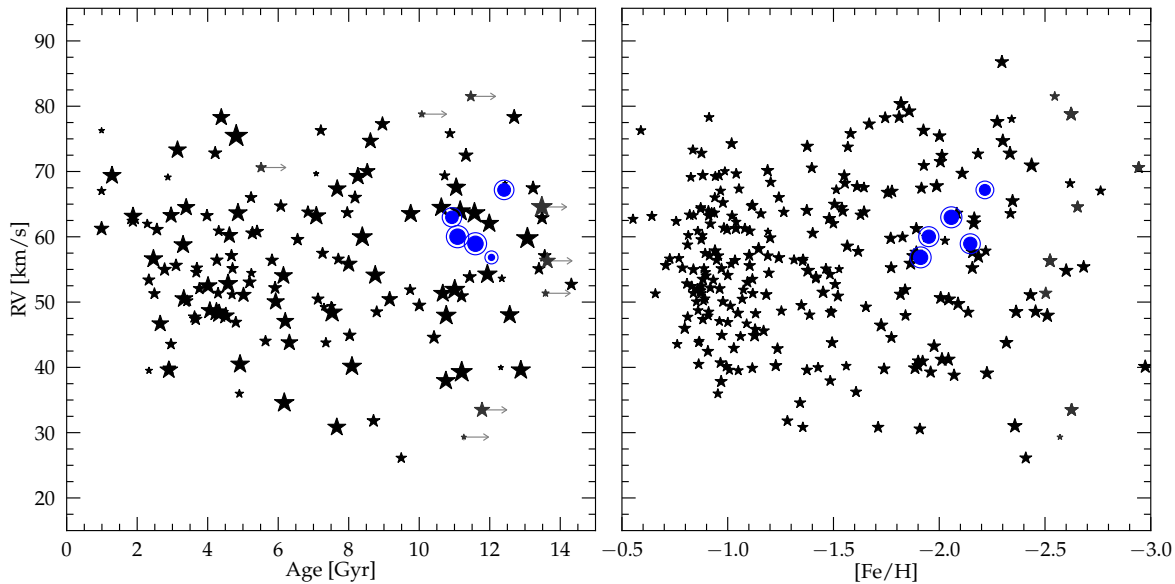}
\caption{\textbf{Left}: Individual radial velocities as a function of age. Symbols with a right-handed arrow show stars which fell out of the lower metallicity range of the isochrones and therefore only lower age limits could be determined. \textbf{Right}: Radial velocities as a function of $\mathrm{[Fe/H]}$. In both panels, an increase of the velocity dispersion with decreasing $\mathrm{[Fe/H]}$ or age is visible. The age-RV distribution becomes non-Gaussian for stars older than $\sim 8$\,Gyr. In the right panel, we plot all stars with a valid CaT-measurement, whereas in the left panel we restrict our selection to the same sample shown in Figure\,\ref{fig_age_fe_relation}. As before, larger symbols indicate stars with higher photometric and chemical precision and blue symbols highlight stars associated with the GCs H2 and H5.}
\label{fig_age_dynamical_bimodality_1}
\end{center}
\end{figure*}

In Figure\,\ref{fig_age_dynamical_bimodality_1} we show the individual distribution of line-of-sight velocities at different metallicities and ages. When we follow the radial velocities of stars with increasing age (left panel in this Figure), the velocity dispersion not only increases systematically towards older stars, but the stars show a slightly bimodal RV distribution: while stars with ages $\sim 7$\,Gyr and younger have a small velocity dispersion around the galactic mean motion, older stars are equally distributed between high ($\sim 70$\,$\mathrm{km\,s^{\rm -1}}$) and low ($\sim 45$\,$\mathrm{km\,s^{\rm -1}}$) RVs.
These observations suggest that the non-Gaussian dynamical pattern of metal-poor stars reported by \citet{Battaglia_06} for the inner regions of the galaxy in fact have a continuation to larger radii.

It is important to note that the existence of a non-Gaussian dynamical distribution of stars only within a specific population bears the risk of introducing a selection bias to any stellar sample, if membership is assigned with a sigma-clipping procedure based on stellar velocities. In such a scenario, preferably members of the population which is not in dynamical equilibrium would be excluded from the sample, in the case of Fornax the oldest and most metal-poor stars.
We therefore re-examine those stars in our sample which we excluded from the analysis due to deviant velocities. From 11 candidates with CaT-metallicities, 9 have $\mathrm{[Fe/H]} \leq -1.4$, and by that fall, e.g., in the metallicity-range of the existing GC systems in Fornax. However, these metallicities also show the expected $\mathrm{[Fe/H]}$-pattern of Halo foreground contaminants (e.g., \citealt{Schroeck_09}, \citealt{Ryan_91}).

Note also that the determination of $\sigma_{sys}$ assumes a (Gaussian) distribution of velocities of stars in dynamical equilibrium. If significant fractions of a stellar system show kinematical substructures (\citealt{McConnachie_07}) or bimodalities (\citealt{Battaglia_06}), as it may be the case for Fornax, the use of the projected velocity dispersion profile to interpret the dynamical state of galaxy field stars in Fornax, e.g. for mass estimations, is ambiguous.

\subsection{Discussion of Age Uncertainties}
\label{section_age_uncertainties}
Parameters that contribute to the age uncertainty ($\sigma age$) of our targets are the photometric errors (in color and magnitude), the uncertainty in metallicity and $\mathrm{[\alpha/Fe]}$, which define the set of isochrones used for each star, as well as uncertainties in the distance modulus and interstellar reddening. In our case, \emph{all} of the above mentioned sources have a significant impact on the total precision and accuracy with which individual stellar ages can be derived. However, absolute ages also depend on the chosen set of isochrones and moreover to the uncertainties in the underlying stellar physics. Therefore, the total age uncertainty includes assumptions on mixing length, core-overshooting, mass loss, etc. in the models, which to discuss is beyond the scope of this study.
In the following, we consequently limit our discussion to uncertainties in relative ages and discuss the net-effects of $\sigma mag$, $\sigma color$, $\sigma \mathrm{[Fe/H]}$, and $\sigma \mathrm{[\alpha/Fe]}$ on the age when derived from RGB stars and under the assumption that the adopted model sequences predict the stellar position in a CMD correctly.

The major sources for statistical error in the age determination are the photometric color uncertainty (here $\sigma (B-V)$) as well as the intrinsic error in the assumption of the stellar metallicity ($\sigma \mathrm{[Fe/H]}$) and alpha-abundance ($\sigma \mathrm{[\alpha/Fe]}$). 
Typical photometric uncertainties for stars in our sample with $V\leq19.5$ are $\sigma (B-V) \approx 0.005$ and $\sigma V \approx 0.003$. 
For $\sigma \mathrm{[Fe/H]}$ and $\sigma \mathrm{[\alpha/Fe]}$ typical values are 0.1\,dex and 0.15\,dex, respectively.

The adopted distance modulus has a strong systematic effect on the age determination, especially for stars close to the tip of the RGB.
For Fornax, a variety of standard candles have been used to determine the distance from Cepheids (\citealt{Greco_05}), tip-RGB magnitudes in the optical (\citealt{Buonanno_99}, \citealt{Bersier_00}) and NIR (\citealt{Pietrzynski_09}), as well as from the luminosity of the red clump (\citealt{Bersier_00}, \citealt{Pietrzynski_03}).
While some studies state an intrinsic error of $\sim 0.1$\,dex (corresponding to a true distance error of $\sim 8$\,kpc), one has to take into account that \emph{all} photometric standard candles require empirical calibrations which are dependent on the assumed metallicity and age in the system (e.g., \citealt{Valenti_04}). Because generally GCs are used to calibrate the photometric standard candles, some systematic differences can be expected when applied to the field star population in a galaxy. Consequently, for systems like Fornax, which show a significant intrinsic variation in $\mathrm{[Fe/H]}$ and age, the total uncertainty of the distance modulus is likely larger than the above stated number, even for up-to-date measurements.
Here, we adopt $(m-M)_{0}=20.84\pm0.12$, the most recent distance determination from \citet{Pietrzynski_09}, determined from the tip-RGB magnitude in the NIR. The quoted uncertainty is not only the intrinsic (statistical+systematic) error of their method, but is also a reasonable reflection of the variation between different existing distance values from various studies in the past (see Table 3 in their paper).

Although the line-of-sight foreground reddening in the direction to Fornax is small (see previous Section), the reddening maps from \citet{Schlegel_98} have a zero-point uncertainty of $0.016$ in $E(B-V)$ and a pixel-to-pixel statistical uncertainty of $0.16\times E(B-V)$, which corresponds to $\sigma (B-V)\approx0.005$ in our field-of-view.
These numbers are similar in size compared to $\sigma (B-V)$ in the photometry and therefore add significantly to the overall error in the age analysis.

Taken together, $\sigma color$ is mainly of statistical nature and dominated by a combination of $\sigma (B-V)$ from the photometry and $\sigma E(B-V)$ from the reddening maps. In contrast, $\sigma mag$ has only a small statistical error, and is dominated by the systematic uncertainty in the distance modulus.

In order to quantify age uncertainties for stars in our sample of different [Fe/H] and age, we generate a fine grid of isochrones between $-2.5\leq\mathrm{[Fe/H]}\leq -0.8$ and for ages between $2$ and $15$\,Gyr with steps of 0.1\,dex and 0.5\,Gyr, respectively. Then, for each point in this parameter space, we determine the age difference when we vary $\sigma mag$, $\sigma color$, $\sigma \mathrm{[Fe/H]}$, and $\sigma \mathrm{[\alpha/Fe]}$ according to their estimated values. 
All calculations are performed for a star at a distance to the tip of the RGB of $0.5$\,mag (in our case $V=19.0$\,mag), which is typical for our sample. 

\begin{figure*}[htb]
\begin{center}
\includegraphics[width=1.0\textwidth]{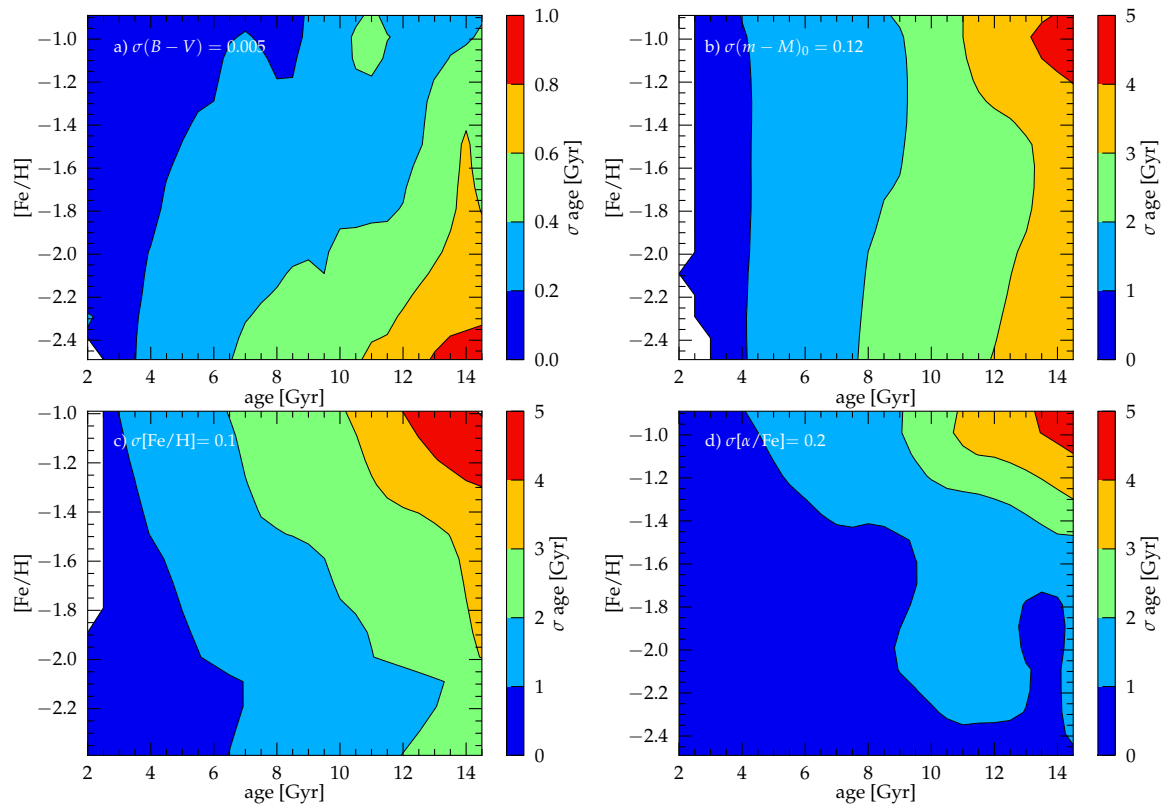}
\caption{Assessment of different sources of uncertainty on our ages: In all cases we evaluate the same grid of [Fe/H] and ages for a star at $V=19.0$. See text for details. \textbf{a)} shows the correlation between $\sigma(V-I)$ and $\sigma age$. Here we choose $\sigma (B-V)=0.005$, a typical value for our photometry. The resulting age uncertainty is generally small ($\leq1$\,Gyr) and increases from young, metal-rich, to old and metal-poor stars. In \textbf{b)} the effect of $\pm0.12$\,mag variation in the distance modulus on the derived stellar ages is shown. For smaller distances generally younger ages are obtained with an increasing effect for older stars. In \textbf{c)} the sensitivity of our age analysis on the assumption of stellar metallicity is shown. Similar to the photometric uncertainty, the effect increases with stellar age but is significantly larger for typical uncertainties of $\sigma\mathrm{[Fe/H]}=0.1$. Finally, \textbf{d)} shows the age sensitivity on the adopted $\mathrm{[\alpha/Fe]}$, relative to a solar ratio which reveals a very similar and only slightly smaller effect on stellar ages than $\sigma\mathrm{[Fe/H]}$.}
\label{fig_age_uncertainties}
\end{center}
\end{figure*}

The results are shown in Figure\,\ref{fig_age_uncertainties}. Generally, we find that ages for older stars become more uncertain. 
As can be seen in panel a), $\sigma (B-V)$ causes $\sigma age \leq 0.5$\,Gyr for the majority of younger and metal-rich populations, which rises to $\sigma age \approx 1$\,Gyr for the oldest stars. From panel c) and d) we find that both $\sigma \mathrm{[Fe/H]}$ and $\sigma \mathrm{[\alpha/Fe]}$ introduce a statistical uncertainty to our ages which varies between $\sigma age \leq 2$\,Gyr for most of our targets up to $\sigma age \approx 5$\,Gyr for old but metal-rich targets. However, since this region is practically not covered by ``real'' stars (see Figure\,\ref{fig_age_fe_relation}), the maximum uncertainty introduced by chemical input parameters should not exceed 3\,Gyr for our stars.
One of the strongest impact on stellar ages comes from the uncertainty in the distance modulus. From panel b), we find that while young stars are accurate to better than 2\,Gyr, old stars become systematically uncertain to as much as 3-4\,Gyr.

Note that the effect of all error-afflicted parameters in the age determination process are highly asymmetric. Specifically the results are more uncertain towards iron-depleted stars.
However, to minimize the cases where a stellar age falls outside the grid of available isochrones, and therefore to maximize the available parameter space in our grid, in Figure ,\ref{fig_age_uncertainties} we evaluate only the difference between the actual and the younger age, and the errors needs to be multiplied by a factor of $\sim1.5$ to obtain the error towards older ages.

In summary, for most of our young and metal-rich targets $\sigma (B-V)$, $\sigma \mathrm{[Fe/H]}$, and $\sigma \mathrm{[\alpha/Fe]}$ will cause an error in age of less than $0.2$, $1$, and $1$\,Gyr, respectively, and consequently a total statistical age uncertainty of $\sigma age \approx 1.5$\,Gyr can be expected in our analysis, topped with a possible systematic shift of at least $1$\,Gyr on the age. For the oldest and metal-poor stars we obtain a statistical error of $\sim2.5$\,Gyr and a systematic accuracy $\sim2$\,Gyr. 
Therefore it is important to stress that, although the age-trends we observe here fall on a fairly well defined sequence, the ages, specifically of old stars, should be interpreted with caution and a sample with both better photometric accuracy as well as a better knowledge about the distance to Fornax is needed to reduce the systematic impact on our results. 

At this point we can revisit the difference in the AMRs between the two distinct pointings in our sample which are shown in Figure\,\ref{fig_age_fe_relation}. First, it should be noted that, although the two fields are centred at slightly different radial positions, the large majority of stars in both fields share a common distance to the center of Fornax. Therefore we can rule out that the bias we observe is caused by any radial variations in the chemical enrichment.
Other than that, the different trend in the ARM could be caused by systematic differences in the CaT-metallicities. We can rule out systematics in [Fe/H] for several reasons: first, the spectra have been taken with the same instrument and have been processed and analyzed with the same pipelines. We do not find a zero-point difference between the MDFs for both fields. Second, there is a subsample of our dataset that was also analyzed in \citet{Battaglia_08}, and we do not observe any systematic offset between the two sets, although it includes stars from both fields. Finally, as can be seen in Figure\,\ref{fig_CaT_calibration}, there is no difference between the two fields when the results are compared to [Fe/H] independently determined from iron absorption lines.
Since we use the same set of isochrones for age-determination, the only systematic source left is the photometric information for our stars, and in fact we see strong evidence that differences in the photometric zero-points are the reason for the putative difference in the chemical enrichment path between Field\,1 and 2.
Due to the large separation of the two fields (almost 1\degree in the sky), the photometry for each of these two subsamples comes from different pointings (see Figure 1 in \citealt{de_Boer_12}). Although the authors used an overlap in the individual frames to find a common photometric zero-point for all fields, a remaining uncertainty of $0.03$\,mag in each filter is typical (see, e.g., \citealt{Coleman_08}) and the zero-point difference in any photometric color can therefore be expected to be different by $\sim0.04$\,mag. This offset is almost an order of magnitude higher than the statistical error evaluated in Figure\,\ref{fig_age_uncertainties} and consequently can cause age differences from 2 to 4\,Gyr, depending on the age and metallicity of the star.
Therefore the offset between the AMRs shown in Figure\,\ref{fig_age_fe_relation} is most likely the result of small zero-point variations in the photometry at different positions in the galaxy. 

\section{The Metallicity Distribution Function}
\label{mdf_chapter}
The MDF gives important insight in the integrated chemical enrichment history of a galaxy and it can help to constrain different enrichment scenarios, especially when it is used in combination with detailed enrichment models (e.g., \citealt{Kirby_11c}, \citealt{Lanfranchi_10}, \citealt{Hendricks_14}). 
Asymmetries or distinct peaks in the MDF can be signs for intense, burst-like star formation or accretion events in the past.
Because dSphs may possibly have contributed to the build-up of the Galactic Halo, the metallicity budgets of dwarf galaxies are also important keys to better understand if, and to what extend these systems have donated their populations to our Galaxy (e.g. \citealt{Helmi_06}, \citealt{Starkenburg_10}). 

Fornax displays a significant radial metallicity gradient, where the metal-rich (and younger) stars tend to be found closer the galactic center (\citealt{Stetson_98}, \citealt{Battaglia_06}).
Consequently, the MDF is a function of radius, if not generally a function of position in the galaxy and it seems necessary to shed light on the \emph{differences} within the MDF at different radii to understand the evolution of a dSph galaxy as a whole.

\subsection{Distinct populations in Fornax?} 

\begin{figure}[htb]
\begin{center}
\includegraphics[width=0.5\textwidth]{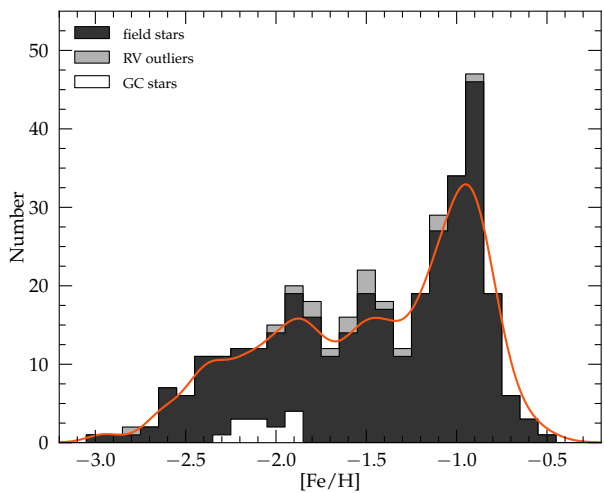}
\caption{MDF from our sample of stars in two outer fields at $r_{ell}\sim~0.6$\degree in the Fornax dSph. Different shadings highlight different components in our sample, where we explicitly highlight the stars which we removed in the process of the radial-velocity clipping. Note, that this selection process did not introduce any bias to our sample, as may be expected from a non-Gaussian underlying velocity distributions. The red line is the error-convolved version of the underlying field-star histogram.}
\label{fig_MDF_1}
\end{center}
\end{figure}

The MDF drawn from our complete sample of 340 field stars is shown in Figure\,\ref{fig_MDF_1}. The binning size is chosen according to the median uncertainty $\sigma \mathrm{[Fe/H]}=0.10$\,dex for our CaT-metallicities. To remove possible binning-biases, we also show the error-convolved interpretation of the same distribution. In addition to the field stars, we also plot the metallicities of stars which we have previously identified as GC members (see Section\,\ref{GCs}) as well as the small sample of stars which fall outside of our RV-membership criterion (see Section\,\ref{radial_velocities}), but may nonetheless be members of the galaxy with non-Gaussian dynamics. In the following we will use the term \emph{population} only for stars who share the same age, while stars with same metallicity will be denoted as \emph{group}, motivated by the fact that such groups can in fact host several generations of stars. 

Generally, the MDF is dominated by a prominent metal-rich group at $\mathrm{[Fe/H]}\approx -1.0$ and a significant fraction of stars at lower metallicities, down to -3.0\,dex. The mean metallicity of our sample is $\mathrm{[Fe/H]}=-1.48$ at a mean radius of $r_{ell}=0.57$\degree. The sample displays two more spikes in the metallicity distribution, one peaking at $-1.9$\,dex, indicating a metal-poor group, and a second one located at $\mathrm{[Fe/H]}\approx-1.4$.

The small group of GC stars coincides with the metal-poor peak in the field-star MDF. Therefore it is possible that this overdensity is the result of accreted GC stars during earlier evolution. Such a scenario is discussed in \citet{Larsen_12a}, who estimate the total fraction of former GC stars in the field star population of Fornax to be $\geq 20\%$. The ``contamination'' of a galaxy by a significant fraction of GC stars might also be the case for other dwarf galaxies (\citealt{Larsen_14}), and should be considered when the MDF is used to interpret the chemical evolution history of these systems.

The peak at $\approx-1.0$\,dex has been observed in all previous studies in more central areas (\citealt{Pont_04}, \citealt{Battaglia_06}, and \citealt{Letarte_10}), and we still find it to be the prominent feature at larger radii. However, this population practically disappears for $r_{ell}\geq0.65$ (see Section\,\ref{radial_gradients}). \citet{Battaglia_06} also find a significant group of more metal-poor stars (which they define by $\mathrm{[Fe/H]}\leq-1.3$) and a very metal-rich group at $\mathrm{[Fe/H]}\approx-0.6$. In a later study, \citet{Amorisco_12} could show that there are indeed distinct dynamical properties amongst the three subgroups of stars.
Our MDF suggests that the ``metal-poor'' group shows an additional peak at $\mathrm{[Fe/H]}=-1.4$ and therefore may in fact be composed of several distinct populations. Although the number of stars in our analysis is still too small to rule out an artifact from small-number statistics, it is remarkable that \emph{all} previously mentioned studies display a peak or a bump in the MDF at $\mathrm{[Fe/H]}\approx-1.4$, which supports our findings (see also \citealt{Coleman_08} who also find three peaks at $\mathrm{[Fe/H]}=-1.0$, $-1.5$, and $-2.0$).

When we use the Kaye's Mixture Modeling (\emph{KMM}) algorithm of \citet{Ashman_94} to divide the sample in several Gaussian-shaped metallicity populations, we find that at least 4 populations are required to obtain an adequate fit to the sample. Importantly, the metal-rich peak is extremely well resembled with a single Gaussian fit. In contrast, it is likely that at least the metal-poor populations ($\leq-1.8$\,dex) should not be described by a Gaussian profile and we consider these stars to be part of one or several non-Gaussian groups. 
With these assumptions, we find that the [Fe/H]-groups in our sample peak at $-0.97$\,dex, $-1.45$\,dex, and $-2.11$\,dex, with relative contributions to the overall sample of respectively 45\%, 18\%, and 37\%. The position of the latter group is determined from the weighted combination of the two most metal-poor Gaussian fits.

\subsection{Comparison between Fornax and Sculptor}

Fornax is about ten times more massive than Sculptor today, but both systems share a similar early chemical enrichment history drawn from their $\alpha$-element evolution (\citealt{Hendricks_14}). When their SFHs are compared, the major difference between them seem to be that Sculptor stopped forming stars $\sim 7$\,Gyr ago (\citealt{de_Boer_12a}), while Fornax kept forming stars almost until today (\citealt{Stetson_98}, \citealt{de_Boer_12}), while both systems have stellar populations as old or older than 12\,Gyr.

The metal-rich peak in the MDF of Fornax as drawn from our sample is strikingly narrow, and in fact can be fitted with a single Gaussian using a FWHM corresponding to the statistical uncertainty of our CaT-metallicities and thus indicating a very low intrinsic metallicity scatter within the group. From the age analysis in Section\,\ref{age_chapter}, we know that this metal-rich group does not consist of a single population, but is in fact the conglomeration of a large, young population with $\sim4$\,Gyr in age and other stars between 4 and 8\,Gyr. Consequently, all stellar populations in Fornax which formed after the end of SF in Sculptor are combined in this metal-rich group of stars.
 
\begin{figure}[htb]
\begin{center}
\includegraphics[width=0.5\textwidth]{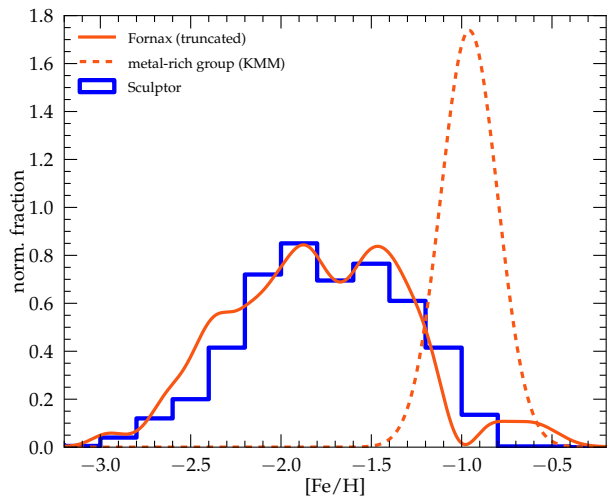}
\caption{Fornax' and Scuptor's MDF show striking similarity when the metal-rich population in Fornax is fitted with a Gaussian and removed from the convolved distribution. This procedure also reveals the existence of a small population at $\mathrm{[Fe/H]}=-0.65$, which previously has been detected in the inner regions of Fornax and displays distinct dynamical properties from the more metal-poor populations (\citealt{Amorisco_12}). Note that Fornax' MDF (solid orange line) and the subtracted peak (dashed orange line) have been rescaled such that the \emph{truncated} MDF resembles the same total area as Sculptor's MDF.}
\label{fig_MDF_sculptor}
\end{center}
\end{figure}

It is possible that this late, intense burst of SF in Fornax was either caused by a late merger event or an otherwise triggered event of SF, e.g., through re-accretion of previously heated or expelled gas as discussed in \citet{Ruiz_13} or \citet{dErcole_99}, or triggered by environmental influences like tidal interactions or ram pressure shocks. In order to mimic a SFH in Fornax which \emph{lacked} such an event and any SF younger than $\sim8$\,Gyr, we use the previously determined \emph{KMM} parameters to fit the prominent, narrow peak with a single Gaussian and then simply subtract this group from the convolved histogram of the full sample (see Figure\,\ref{fig_MDF_sculptor}).

The metallicity distribution of Sculptor is adopted from the recent study of \citet{Romano_13}. Specifically, it has been derived from the centrally constrained sample of \citet{Kirby_08} and the sample from \citet{Battaglia_08} that provides the metallicity distribution of stars in the outer parts. Therefore Sculptor's MDF represents a balanced distribution of stars from the central area to the very outskirts of the galaxy.

Remarkably, the truncated MDF, rescaled to the now smaller stellar budget, exactly matches the MDF of Sculptor. 
The small group at $\mathrm{[Fe/H]}\approx-0.7$ visible in the truncated MDF of Fornax has been previously identified in more central regions of this galaxy (e.g \citealt{ Amorisco_12}) and is likely composed of stars younger than 2\,Gyr.
Therefore, the distribution of metals in Fornax and Sculptor becomes identical at exactly the point when the additional SFH of Fornax with respect to Sculptor is removed from the sample. Consequently, it is likely that these two systems have a very similar enrichment history before Sculptor stopped forming stars 7\,Gyr ago. Such a synchronous evolution between the two galaxies would be in good agreement with their similar $\mathrm{\alpha}$-element evolution (\citealt{Hendricks_14}), which requires a similar chemical enrichment at least during the first $\sim1$Gyr. 

Vice versa to a scenario in which Fornax late SF is caused by a late accretion of additional material, the difference in the late evolution of the two galaxies can possibly be evoked by their individual ability to \emph{retain} their reservoir of gas. Starting with a similar initial mass, the observed properties in their MDF and in the evolution of $\mathrm{\alpha}$-elements would be evoked if Sculptor lost a significant fraction of its gas through tidal stirring, while Fornax did not. Such a scenario may be supported by the orbital properties of both galaxies, because Sculptor's estimated perigalactic distance ( about $68$\,kpc) is by a factor of $\sim2$ smaller than that of Fornax (\citealt{Piatek_06}), while it is likely to follow an orbit with higher excentricity. In this case Sculptor did experience stronger (and more frequent) tidal forces which could explain the early loss of gas. Note, however, that orbital parameters for both dSphs come with large uncertainties and it seems not advisable to draw strong conclusions from them.

\section{Radial Gradients}
\label{radial_gradients}

Fornax displays a well known significant radial metallicity gradient, where the metal-rich stars tend to be found closer the galactic center (\citealt{Battaglia_06}). From photometric studies we know that this observation corresponds to an actual age gradient within the galaxy (\citealt{Stetson_98}, \citealt{de_Boer_12}, \citealt{del_Pino_13}). A radial population gradient seem to be a common feature amongst dSphs (e.g., \citealt{Harbeck_01}, \citealt{Leaman_13}), and it is commonly interpreted as a gradual concentration of the remaining gas within a galaxy towards the center of its gravitational potential accompanied with an outside-in SF. Alternatively, radial gradients could be the result of a differential binding energy with galactocentric radius within the galaxy. Internal and environmental gas-removing processes such as SN feedback, tidal interactions or ram-pressure stripping will in this case more easily remove potentially star forming material from the outer parts of the galaxy, while the most centrally located gas exhibits the highest likelihood of being hold in the galaxy and can subsequently serve as birthplace for new generations of stars.

Our sample is focused on the chemodynamical properties of stars primarily in the outer parts of Fornax, and consequently is not well suited to study overall radial trends in this galaxy. However, since the stars are selected between $r_{ell}\sim0.3$-$0.8$\degree, we can observe a general population-trend for different radii.

\begin{figure}[htb]
\begin{center}
\includegraphics[width=0.5\textwidth]{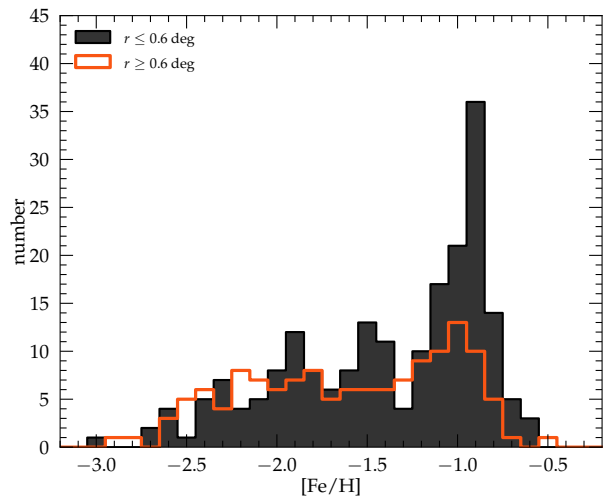}
\caption{Radial variations in the MDF of Fornax. Stars with $r_{ell}\leq 0.6$\degree (black) are clearly dominated by the metal-rich peak at $\mathrm{[Fe/H]}=-0.97$ and two other subgroups are visible. The MDF outside this interval (red) indicates an almost equal distribution between stars of all metallicities.}
\label{fig_MDF_gradient}
\end{center}
\end{figure}

Figure\,\ref{fig_MDF_gradient} shows the MDF when we separate the stars between those who have a galactocentric distance smaller than $0.6$\degree, and those located at larger radii.
It is clearly visible, that the more centrally concentrated stars have a more metal-rich distribution than the stars in the outermost areas. 
In the inner MDF all three peaks we discussed before are present, and the metal-rich group is the dominant feature. In contrast, when only the outer stars are examined, the
peak at $\mathrm{[Fe/H]}\approx -1.0$ is barely visible (and will disappear for $r_{ell}\geq0.65$\degree), and we find an even distribution of stars over the entire metallicity range.

\begin{figure}[htb]
\begin{center}
\includegraphics[width=0.5\textwidth]{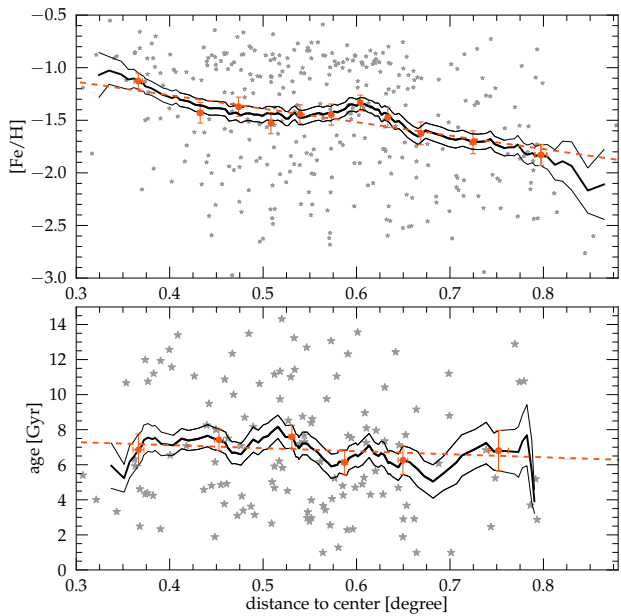}
\caption{\textbf{Top:} Radial metallicity gradient in the our sample. The floating mean [Fe/H] (thick black line) shows a clear decrease of $\mathrm{[Fe/H]}$ with distance to the galactic center. The thin black lines indicate the 1-$\sigma$ uncertainty interval. Red symbols show the same trend from independent subsamples of $\sim30$ stars and the red dashed line shows the best linear fit to this trend. The data is in reasonable agreement for a linear slope, for which we find best fitting parameters of $-1.28\pm0.25$\,dex/degree, corresponding to $-0.50\pm 0.10$\,dex/kpc. Small, gray symbols show the actual distribution of star in our sample.
\textbf{Bottom:} No significant radial age trend in the our sample. When a linear function is fitted to the data, we find a slope of $-1.73\pm1.96$\,Gyr/degree, and consequently no significant age gradient with distance to the galactic center within the radial range of our sample.}
\label{fig_radial_trends}
\end{center}
\end{figure}

To investigate the detailed trend of $\mathrm{[Fe/H]}$ with $r_{ell}$, we first sort the stars in our sample with increasing distance to the centre an then compute a floating mean for a subsample of 80 stars. The result is shown in the top panel of Figure\,\ref{fig_radial_trends}. We find that the mean metallicity drops steadily from $\mathrm{[Fe/H]}\sim-1.2$\,dex at 0.4\degree to $-1.8$\,dex for stars at 0.8\degree. 
The bump in the distribution with a steeper trend for $r_{ell}\geq0.65$\degree is present in \emph{both} fields independently, which indicates that this is a real feature caused most likely by a clearly defined upper radial boundary for stars belonging to the metal-rich group (and consequently with ages younger than $\sim8$\,Gyr).
When we approximate the radial decline of $\mathrm{[Fe/H]}$ with a linear function, we find a slope of $-1.28\pm0.25$\,dex/degree, corresponding to $-0.50\pm 0.10$\,dex/kpc (and $-0.37\pm 0.07$\,dex/$r_{c}$) when we assume $r_{c} = 0.293$\degree and $d=147$\,kpc. 

We can perform a similar analysis for the radial distribution of stellar ages, shown in the bottom panel of Figure\,\ref{fig_radial_trends}. Here we cannot make use of the full sample of stars with CaT-metallicities but have to restrict our analysis to the same selection of stars we presented in Section\,\ref{age_chapter}. Interestingly, we do not find a significant trend of the mean stellar age with galactocentric distance. The mean correlation appears flat with a slope of $-1.73\pm1.96$\,Gyr/degree.

In addition to the statistical uncertainties, it is possible that we introduce a systematic bias in the radial age trend, since the youngest stars (which are most abundant at small radii) are not fully sampled when stars along the tip of the RGB are investigated. Furthermore, the photometric error of stars causes \emph{only} members from the oldest and youngest populations to be discarded in the analysis because only they can \emph{artificially} fall outside of the isochrone range due to their statistic and systematic uncertainties. Therefore, it is hard to draw conclusions about the quantitative distribution of stellar ages from spectroscopic samples like ours, because such analysis requires a large statistical sample with no selection bias and a negligible number of systematic outliers like AGB interlopers or incorrect alpha-assumptions for individual stars. 

\section{Summary}

We have presented precise radial velocities and CaT-metallicities for a large sample of 340 stars in two distinct outer fields of the Fornax dSph from intermediate resolution spectra ($R\sim16,000$).
While the inner regions of Fornax have been studied in detail by, e.g., \citet{Pont_04}, \citet{Battaglia_06}, or \citet{Letarte_10}, the outer region of this galaxy have not been studied systematically although it is known that Fornax -- like dSphs in general -- displays a strong variation in its chemodynamical properties with distance to the galactic centre (\citealt{Battaglia_06}). The present analysis is intended to fill this gap and should help, in combination with the existing spectroscopic samples in Fornax, to decipher its evolution and reduce the level of selection bias for important galactic parameters. 
In the following we summarize our detailed results.

\begin{itemize}

\item Stars in our sample show a wide range in metallicity, between $\mathrm{[Fe/H]}=-0.5$ and $-3.0$\,dex. The MDF in the outer fields of Fornax is dominated by a distinct metal-rich group of stars at $\mathrm{[Fe/H]}=-0.97$, which is seen out to $\sim0.65$\degree, where it disappears abruptly. In total we observe three subgroups at $\mathrm{[Fe/H]}=-0.97$, $-1.45$, and $-2.11$ with a relative contribution to the total sample of 45\%, 18\%, and 37\%, respectively. Our sample contains $75$ stars with $\mathrm{[Fe/H]}\leq-2.0$, increasing the database of rare metal-poor stars in Fornax by a factor of two.

\item When we remove the most metal-rich (and youngest) populations from our sample, the truncated MDF becomes identical to the one observed in Sculptor. This striking similarity is a strong evidence that these systems evolved identically at early times, with the only difference that Fornax experienced a late and intense episode of SF, which could have been triggered, e.g., through a merger event (\citealt{Yozin_12}), by re-accretion of previously expelled gas (\citealt{Ruiz_13}, \citealt{dErcole_99}), or by environmental influences such as tidal interactions.

\item Our data confirms a radial gradient of $\mathrm{[Fe/H]}$ with galactocentric distance in Fornax. The gradient within our radial coverage reasonably resembles a linear slope with $-1.28\pm0.25$\,dex/degree.
In contrast, we do not observe a significant age gradient for increasing radii for which a similar fit yields $-1.73\pm1.96$\,Gyr/degree.
However, the age-gradient should be interpreted with caution since the available sample is small and the analysis is likely to suffer from systematic selection biases.

\item We combine the independent $\mathrm{[Fe/H]}$ measurements from individual iron lines in our spectra measured with \textit{SPACE} (\citealt{Boeche_13}) to test different CaT-calibrations over more than 2\,dex in $\mathrm{[Fe/H]}$. We find best agreement with the calibration equations provided in \citet{Carrera_13}, while classical GC-calibrations yield CaT-metallicities systematically too large for stars below $\mathrm{[Fe/H]}\approx-1.8$. 
Finally, we identify the actual approach of fitting the CaT absorption features as a possible source of systematic offsets in the calibration, leading to systematic shifts between different datasets as large as $0.5$\,dex in [Fe/H]. To avoid such effects, the same line-fitting technique should be applied to the stars, as it has been used to derive the calibration equations of choice.

\item From our sample we can associate nine stars to the GC H2 and three stars to H5. We find $\mathrm{[Fe/H]}=-2.04\pm0.04$ and $RV=59.36\pm0.31$\,$\mathrm{km\,s^{\rm -1}}$ for H2 and $\mathrm{[Fe/H]}=-2.02\pm0.11$ and $RV=59.39\pm0.44$\,$\mathrm{km\,s^{\rm -1}}$ for H5, in excellent agreement with previous findings. In the case of H2, we provide the largest sample of individual measurements for RV and $\mathrm{[Fe/H]}$, as previously only three individual members have been examined by \citet{Letarte_06}.

\item We combine our information about $\mathrm{[Fe/H]}$ with $\mathrm{[\alpha/Fe]}$ for individual stars to derive ages for these targets with object-specific isochrones. The general trend in the AMR indicates a chemical enrichment in three-phases: 
a steep increase in $\mathrm{[Fe/H]}$ at early ages, followed by a significantly slower and almost flat enrichment for stars younger than $\sim8$\,Gyr. 
Finally, the AMR show signs for a second, fast enrichment in metallicity during the last $3$\,Gyr, subsequent to a strong, young stellar population in the galaxy.
This picture shows, that the dominant metal-rich population in this galaxy does not reflect a single stellar population, but instead contains stars of at least $5$\,Gyr in age during an evolutionary phase where the ISM was not significantly enriched.
These observations are in good agreement with predictions from our earlier chemical evolutionary model where the SF efficiency increases with time in a series of bursts -- a scenario which also possibly explains the peculiar, metal-poor position of the knee in the evolution of [Mg/Fe] (see \citealt{Hendricks_14}). 
However, the model does not predict a significant fraction of young stars with ages $3-5$\,Gyr, nor the subsequent steep increase in [Fe/H] which we observe in the data. Therefore, we tentatively propose that this younger population(s) may have been the result of an externally triggered SF episode, for which a simple leaky-box model does not account for.

\item Our evolutionary scenario agrees well with the empirical SFH from \citet{Weisz_14}, except in two aspects:
First, our model predicts a larger fraction of old stars. This difference can be explained with a radial SFH gradient in Fornax observed in \citet{de_Boer_12}.
Second, our model does neither predict a significant fraction of stars younger than $\sim5$\,Gyr, nor a significant fraction of SF during that time. This difference can be explained with
an environmentally-triggered episode of SF which a leaky-box model cannot take into account.

\item The few individual GC stars in our sample fall on top of the field-star AMR relation within the respective uncertainties. This implies a similar chemical enrichment of the protostellar material out of which both GC and field stars formed. However, given the large uncertainty on ages -- especially for old, metal-poor stars -- we cannot rule out the possibility of small or moderate differences in the chemical enrichment between the two environments.

\item A detailed analysis of several sources of statistical and systematic uncertainties in the age determination shows that the total error is dominated by systematic effects.
Hereby, the uncertainty in the distance modulus pose a major source of error which can alter the derived ages of stars by several Gyr depending on their intrinsic age and metallicity. Furthermore, we identify small zero-point variations between photometric frames as an additional source of systematic error, which in our case evokes an apparently different chemical enrichment pattern for the two galactic areas we investigate. Given the high sensitivity of ages on photometric parameters, a star-by-star reddening correction becomes necessary, even for galaxies with very low line-of-sight interstellar extinction.

\item When we combine the dynamical and chemical information of our sample, we find that different populations also display complex dynamical properties, which has been previously observed for the inner regions of Fornax (\citealt{Battaglia_06}, \citealt{Amorisco_12}). Specifically, the velocity dispersion continuously increases from $\sigma_{sys}\approx 7.5\,\mathrm{kms^{-1}}$ to $\geq14\,\mathrm{kms^{-1}}$ from the highest to the lowest metallicities. This signature, together with the observed radial metallicity trend in Fornax, supports an outside-in SFH in a Dark Matter dominated halo, under the assumption that subsequent bursts of SF generally increase in [Fe/H].
However, the large velocity dispersion at low metallicities could alternatively be the result of a non-Gaussian velocity distribution amongst stars older than $\sim8$\,Gyr, with a flat distribution of RVs between 40 and 70 $\mathrm{km\,s^{\rm -1}}$. 
If these complex dynamical signatures are real and not caused by small number statistics or local inhomogeneities within the galaxy, they can be a sign for accreted stellar systems. 

\item Finally, we do not observe significant differences in the chemical and dynamical properties between the two distinct fields of our survey at the same radii but opposite sides, which suggests that there are only small \emph{random} local variations or inhomogeneities within the galaxy.

\end{itemize}

This work confirms that there are significant differences between the inner regions of dSphs and their outer parts, where some details can only be revealed if a statistically large number of stars is available at different radial positions in a galaxy. Our sample, although large, still lies well within the tidal radius of the galaxy (with a significant fraction of stars only out to $r_{ell}=0.75\degree$). However, the tidal radius of Fornax reaches out to $\geq1$\degree, and \citet{Battaglia_06} found that Fornax likely has a significant number of members beyond that radius. In order to get a full picture of Fornax' history, it is important to add to the existing stars for which we already have basic chemodynamic information, a sufficiently large sample of stars taken from the \emph{real} periphery of the galaxy, at radii $\geq0.8$\degree and beyond the tidal radius. The same necessity applies to other dSphs, for which the available -- photometric and spectroscopic --  samples are, in most cases, even stronger biased towards the center of these objects.

\begin{acknowledgements}
B. Hendricks thanks G. Battaglia for providing the extended catalog of EWs of Fornax field stars, and T. de Boer for the tables with photometric SFRs. We thank G. Battaglia and an anonymous referee for a careful reading of the manuscript and helpful comments which improved the content of this paper. BH and AK acknowledge the German Research Foundation (DFG) for funding from Emmy-Noether grant Ko 4161/1. This work was in part supported by Sonderforschungsbereich SFB 881 "The Milky Way System" (subproject A5) of the DFG . C.I.J. acknowledges support through the Clay Fellowship administered by the Smithsonian Astrophysical Observatory. This work was partly supported by the European Union FP7 programme through ERC grant number 320360.
\end{acknowledgements}

\begin{appendix}
\section{Chemodynamical parameters for field stars and GC members in Fornax}

\begin{table*}[htb]
\centering
\caption{Chemodynamical parameters for field stars and GC members in Fornax  --  astrometry, photometry, velocities}
\begin{tabular}{lccccccccc}
\hline
\hline

Star ID	 & $\alpha$	 	 & $\delta$	 	 & S/N	 	 & RV [$\mathrm{km\,s^{\rm -1}}$]	 & $\sigma$ RV [$\mathrm{km\,s^{\rm -1}}$] & $V$	 & $\sigma V$	 & $V-I$	 & $\sigma (V-I)$ \\\hline
    2	 &   2h37m12.20s	 & -34d51m50.18s	 &  20.7	 & 47.95	 &  0.97	 & 18.938	 &  0.031	 &  1.179	 &  0.039	  \\
    3	 &   2h37m14.11s	 & -34d51m30.82s	 &  18.5	 & 39.58	 &  1.22	 & 19.094	 &  0.041	 &  1.301	 &  0.048	  \\
    4	 &   2h37m16.63s	 & -34d50m11.18s	 &  17.7	 & 46.48	 &  0.90	 & 19.277	 &  0.042	 &  1.069	 &  0.049	  \\
    5	 &   2h37m19.31s	 & -34d43m49.19s	 &  22.6	 & 26.40	 &  1.32	 & 19.457	 &  0.052	 &  1.211	 &  0.059	  \\
    6	 &   2h37m22.88s	 & -34d53m40.31s	 &  15.2	 & 56.62	 &  0.91	 & 19.378	 &  0.051	 &  1.272	 &  0.056	  \\
    7	 &   2h37m25.62s	 & -34d42m43.70s	 &  29.3	 & 50.91	 &  0.68	 & 19.092	 &  0.029	 &  1.283	 &  0.035	  \\\hline

\end{tabular}
\tablefoot{Astrometry for all our targets is calibrated to the USNO-1B system. For detailed information see Walker et al. (2006). Tables\,\ref{table_field_stars_1}, \ref{table_field_stars_2}, and \ref{table_field_stars_3} list all stars which passed the selection process described in Section\,\ref{chap_02}.} The entire table is published in the electronic version of the journal, and a portion is shown here to guide the reader to the content.
\label{table_field_stars_1}    
\end{table*}

\begin{table*}[htb]
\centering
\caption{Chemodynamical parameters for field stars and GC members in Fornax  --  CaT-metallicities, ages, EWs}
\begin{tabular}{lccccccc}
\hline
\hline

Star ID	 & $\mathrm{[Fe/H]_{CaT}}$	 & $\sigma \mathrm{[Fe/H]_{CaT}}$	 & age [Gyr]	 & $\mathrm{EW_{CaT1}}$	 & $\mathrm{EW_{CaT2}}$	 & $\mathrm{EW_{CaT3}}$	 & $\mathrm{EW_{Mg\,I}}$  \\\hline
    2	 & -1.84	 &  0.10	 & 10.8	 &  0.993	 &  2.363	 &  1.984	 &  0.231  \\
    3	 & -1.37	 &  0.10	 & 12.9	 &  1.097	 &  2.820	 &  2.434	 &  0.351  \\
    4	 & -2.07	 &  0.11	 & $-$	 &  0.730	 &  1.954	 &  1.719	 &  0.211  \\
    5	 & -1.33	 &  0.09	 & 13.0	 &  1.139	 &  3.027	 &  2.097	 &  0.356  \\
    6	 & -0.73	 &  0.10	 &  2.5	 &  1.397	 &  3.521	 &  2.921	 &  0.362  \\
    7	 & -1.52	 &  0.09	 & 11.2	 &  1.092	 &  2.803	 &  2.138	 &  0.320  \\\hline

\end{tabular}
\tablefoot{$\mathrm{[Fe/H]_{CaT}}$ refers to metallicities determined from the CaT. Column\,8 gives the EW of the Mg\,I line at 8806.8\,\AA. The entire table is published in the electronic version of the journal.}
\label{table_field_stars_2}    

\end{table*}

\begin{table*}[htb]
\centering
\caption{Chemodynamical parameters for field stars and GC members in Fornax  --  Fe-metallicities, $\mathrm{\alpha}$-abundances}
\begin{tabular}{lccccccccc}
\hline
\hline

Star ID	 &  $\mathrm{[Fe/H]_{HR}}$	 & $\sigma \mathrm{[Fe/H]_{HR}}$	 & [Mg/Fe]	 & $\sigma$ [Mg/Fe]	 & [Si/Fe]	 & $\sigma$ [Si/Fe]	 & [Ti/Fe]	 & $\sigma$ [Ti/Fe] & $\chi^{2}$	  \\\hline

    2	 & -1.57	 &  0.04	 &  $-$	 &  $-$	 &  $-$	 &  $-$	 &  0.00	 &  0.05	& 1.90	  \\
    3	 & -1.39	 &  0.04	 &  $-$	 &  $-$	 &  0.02	 &  0.06	 &  0.39	 &  0.05	& 2.00	  \\
    4	 & -2.22	 &  0.05	 &  $-$	 &  $-$	 &  $-$	 &  $-$	 &  $-$	 &  $-$	& 2.20	  \\
    5	 & -1.36	 &  0.04	 &  $-$	 &  $-$	 &  $-$	 &  $-$	 &  $-$	 &  $-$	& 2.47	  \\
    6	 & -0.95	 &  0.03	 &  $-$	 &  $-$	 &  $-$	 &  $-$	 &  0.09	 &  0.04	& 1.70	  \\
    7	 & -1.51	 &  0.04	 &  $-$	 &  $-$	 &  $-$	 &  $-$	 & -0.02	 &  0.05	& 2.60	  \\\hline
\end{tabular}
\tablefoot{$\mathrm{[Fe/H]_{HR}}$ are the results derived from individual Fe absorption lines. The uncertainties for individual abundances give the statistical error discussed in Hendricks et al. (2014) and Column\,10 indicates the quality of the spectral fit in terms of a $\chi^2$ value.}
\label{table_field_stars_3}    

\end{table*}

\end{appendix}

\end{document}